\documentclass[a4paper,fleqn]{cas-sc}

\usepackage[authoryear,longnamesfirst]{natbib}

\usepackage{times}

\usepackage{mdframed}
\usepackage{balance}
\usepackage[utf8]{inputenc}
\usepackage[small]{caption}

\usepackage{booktabs} 
\usepackage[ruled]{algorithm2e} 

\usepackage[utf8]{inputenc}
\usepackage[T1]{fontenc}   
\usepackage{url}
\usepackage{booktabs}       
\usepackage{microtype}
\usepackage{xcolor}
\usepackage{graphics} 

\usepackage{amssymb}
\usepackage{amsmath,amsfonts,amstext,amsthm}
\usepackage{thmtools} 
\usepackage{mathtools} 

\usepackage{natbib}
\usepackage[inline]{enumitem}
\setlist{leftmargin=3mm}
\usepackage{booktabs, makecell}
\usepackage{multirow}
\usepackage{cleveref}
\Crefname{equation}{Eq.}{Eqs.}

\usepackage{pgfplots}
\usepackage{tikz}
\usetikzlibrary{calligraphy}
\usepackage[switch]{lineno}

\renewcommand{\cite}[1]{\citep{#1}}

\DeclareMathOperator*{\argmax}{argmax}

\usepackage{pgfplots}
\usepackage{pgfplotstable}

\usepackage{tikz}
\usetikzlibrary{shapes.misc, automata, positioning, arrows, arrows.meta}

\usepackage{breakcites}

\usepackage{thm-restate}

\newtheorem{lemma_}{Lemma}[section] 
\newtheorem{lemma}[lemma_]{Lemma}

\newtheorem{theorem}[lemma]{Theorem}

\newtheorem{observation}[lemma]{Observation}

\theoremstyle{definition}
\newtheorem{definition}[lemma]{Definition}
\newtheorem{example}{Example}

\SetAlCapSkip{.5em}

\DeclareMathOperator{\Pro}{\mathbb{P}}

\newsavebox{\foobox}

\DeclareMathOperator{\poly}{poly}
\renewcommand{\preceq}{\preccurlyeq}
\newcommand{\proper}{+}

\newcommand{\given}{{\,|\,}}

\def\tsc#1{\csdef{#1}{\textsc{\lowercase{#1}}\xspace}}
\tsc{WGM}
\tsc{QE}
\tsc{EP}
\tsc{PMS}
\tsc{BEC}
\tsc{DE}

\begin{document}
\let\WriteBookmarks\relax
\def\floatpagepagefraction{1}
\def\textpagefraction{.001}
\shorttitle{Persuading Stable Matching}
\shortauthors{J. Shaki et~al.}

\title [mode = title]{Persuading Stable Matching}

\author[1]{Jonathan Shaki}

\cormark[1]


\ead{jonath.shaki@gmail.com}



\affiliation[1]{organization={Bar-Ilan University}}

\author[2]{Jiarui Gan}


\ead{jiarui.gan@cs.ox.ac.uk}



\affiliation[2]{organization={University of Oxford}}

\author[1]{Sarit Kraus}
\ead{sarit@cs.biu.ac.il}

\cortext[1]{Corresponding author}

\begin{abstract}
%
In bipartite matching problems, agents on two sides of a graph want to be paired according to their preferences. The stability of a matching depends on these preferences, which in uncertain environments also reflect agents' beliefs about the underlying state of the world. We investigate how a principal---who observes the true state of the world---can strategically shape these beliefs through Bayesian persuasion to induce stable matching that maximizes a desired utility.
Due to the general intractability of the underlying matching optimization problem as well as the multi-receiver persuasion problem, our main considerations are two important special cases: (1) when agents can be categorized into a small number of types based on their value functions, and (2) when the number of possible world states is small. For each case, we study both public and private signaling settings.
Our results draw a complete complexity landscape: we show that private persuasion remains intractable even when the number of worlds is small, while all other settings admit polynomial-time algorithms. We present efficient algorithms for each tractable case and prove NP-hardness for the intractable ones. These results illuminate the algorithmic frontier of stable matching under information design and clarify when optimal persuasion is computationally feasible.
\end{abstract}

\maketitle

\section{Introduction}

Bayesian persuasion, introduced by \citet{kamenica2011bayesian}, is a framework wherein a persuader seeks to strategically design a signaling scheme that conveys information to influence a decision-maker's choices in the persuader's desired direction \cite{KamenicaSurvey}. 
Rooted in Bayesian game theory, this approach acknowledges the decision maker's rational updating of beliefs and freedom of action, while still leaving a place for the principal to incentivize them.
The stable matching problem studied first by \citet{gale1962college}, involves two sets of agents, such that every agent from one set ranks the agents from the other set by some order. The goal is to find a bijection from one side to the other, such that no two agents, one from each side, prefer each other over their matched partners. It was proven that such a bijection, called a stable matching, always exists, and might not be unique. In such cases, computing a stable matching with the maximum weight \cite{vate1989linear,chen2012maximum} may be of interest, where weights are defined for each possible matched pair.

Both Bayesian persuasion and the weighted stable matching problem assume that agents are free to choose actions.
A principal who cannot directly change the agents' utilities wants to steer the agents' decisions in directions that benefit the principal as much as possible. Given this, we combine these problems into what we call {\em persuading stable matching}. In this setting, we have multiple possible worlds, two sets of agents to be matched in pairs, and a principal who obtains utilities from the matched pairs. The agents' utilities for each other, as well as the principal's utilities for the pairs, all depend on the realized world. 
As in the standard Bayesian persuasion model, the principal observes the realized world, while each agent only holds a probabilistic belief.
The principal commits to a signaling policy, signaling the agents about the realized world to influence their beliefs and, in turn, their preferences about whom they would like to be matched to. 
Such signaling capability provides additional leverage for a matchmaker to expand the set of stable matching.

\begin{table*}[t]
\centering
\renewcommand{\arraystretch}{1.3}
\small
\begin{tabular}{@{\extracolsep{6pt}}lcccc}
\toprule
      & \makecell{C types}  
      & \makecell{C worlds} 
      & \makecell{C worlds \& non-degeneracy} 
      & \makecell{non-degeneracy} \\ 
\midrule
\em public 
    & P {\small (Thm.~\ref{thm:type-model-time-complexity})} 
    & \multirow{2}{*}{\makecell{NP-h {\small (Thm.~\ref{thm:nph})}\\(even for 1 world)}}
    & P {\small (Thm.~\ref{thm:public-small-worlds-poly})}
    & NP-h {\small (Thm.~\ref{thm:arbitrary-non-degenerate-hardness})}
    \\
\em private 
    & P {\small (Thm.~\ref{thm:P-small-types-private})} 
    & 
    & NP-h {\small (Thm.~\ref{thm:nph-small-worlds-private})}
    & NP-h  {\small (implied by Thm.~\ref{thm:nph-small-worlds-private})}
    \\     
\bottomrule\\
\end{tabular}
\caption{Summary of results. (``C'' stands for ``a constant number of''.)}
\label{tab:summary}
\end{table*}

\paragraph{Motivation} 
Our framework applies to stable matching applications where the matchmaker possesses critical information about agents' preferences that the agents themselves do not have, while the matchmaker has its own objectives over how the agents should be matched. 
A typical example is student-to-university or internship matching, where students apply through centralized platforms that match them to institutions based on both sides’ preferences \cite{roth1984evolution, diebold2014course, ha2020college, prakhov2020matching}. This centralized approach has been practiced in many countries \cite{pathak2008leveling, roth1984evolution, chen2006school, kojima2009incentives}. The matchmaker often has access to richer information, such as the full academic records, while universities might only see aggregate indicators like GPA or high school rankings.
Beyond merely coordinating preferences, the principal may also have policy goals, such as promoting diversity \cite{prakhov2020matching} or enhancing overall social welfare.

Another prominent example arises in ride-sharing systems, where a platform (e.g., Uber or Waze) matches passengers with drivers \cite{wang2018stable,delle2023social}. Here, passengers and drivers both have preferences over potential partners, shaped by routes, pricing, or timing considerations. These preferences are influenced by real-time supply/demand and traffic conditions---information known to the platform but not to the agents. This informational asymmetry enables the platform to coordinate matches that are not only stable from the agents’ perspective but also efficient from a global welfare standpoint.
Our framework provides a principled way for a well-informed principal to intervene and guide matching outcomes toward socially desirable or policy-driven goals.

\subsection{Challenges and Contributions}

We are interested in the algorithmic aspects of optimal signaling policy design for the principal, in both public and private persuasion.
To compute optimal policies, the first challenge we face is the complexity barrier inherited from both the weighted stable matching (WSM) problem and the multi-receiver persuasion problem. 
Both can be viewed as simplifications of our problem and are known to be intractable.
Given this, we identify two special cases of our problem that are of sufficient practical significance and may potentially admit efficient algorithms. In one case, the agents can be grouped into a small number of types according to the value functions. In another, the number of possible worlds is small.
We then provide a complete analysis of the computational complexity in these cases.

Our main contributions are as follows:

\begin{itemize}
\item 
{\bf Revelation principle and policy structure.}
We first characterize the structure of optimal signaling policies and establish a revelation principle tailored to the matching setting. 
The revelation principle differs considerably from its counterpart in the standard persuasion model. It reveals the necessity of signaling additional information that encodes the agents' preference orders, besides information signaled in the standard model. This foundational result underpins our subsequent algorithmic developments.

\item 
{\bf Efficient algorithms for few agent types.}
Given the general intractability of the problem, we consider the case with a small number of agent types. We prove the tractability of this case for both public and private persuasion by exposing efficient algorithms to compute optimal policies. 
Key to our approach is identifying a set of {\em useful matchings} via exploiting anonymity in the agents' identities. To this end, we introduce the notion of {\em prototypes} and demonstrate an equivalence between matchings of the same prototype, which allows us the flexibility of substituting matchings signaled by a policy with a small set of useful matchings that are universal irrespective of the agents' beliefs.

\item 
{\bf Few worlds and non-degeneracy.}
We then consider the setting with a small number of possible worlds. For public persuasion, we provide an efficient algorithm under a mild non-degeneracy condition. Notably, the problem remains hard even with a single world in the absence of this assumption. Our non-degeneracy condition parallels that of \cite{xu2020tractability}, but extends it to settings with more than two actions per agent---an extension that turns out non-trivial. We offer a generalized definition along with new algorithmic techniques to solve this setting. Both the conceptual extension and the algorithmic design may be of independent interest for broader classes of public persuasion problems involving multiple agents and utility externalities.

\item 
{\bf Private persuasion: subtypes and hardness.}
Finally, we consider {\em private} persuasion, where signals are sent through private channels to the agents. 
We discover that, in the type-based setting, private signaling must distinguish between agents---even among those of the same type. To capture this, we introduce the concept of {\em subtypes}, which enables us to extend our public persuasion algorithms to the private case. We also establish hardness results for private persuasion even under both non-degeneracy and a small number of worlds. These results rely on a delicate reduction that remains robust under small perturbations of instance parameters, combining insights from fine-grained complexity results on stable matchings with our own novel construction.
\end{itemize}

\noindent \Cref{tab:summary} summarizes our results.

\subsection{Related Work}

Our work is in line with a series of works motivated by recent interests in the algorithmic aspects of various forms of Bayesian persuasion \cite{castiglioni2022bayesian,castiglioni2021persuading,HahnHS20,BabichenkoTZ21,rabinovich2015information,xu2015exploring,Dughmi2017,badanidiyuru2018targeting,castiglioni2020persuading,gan2022bayesian,zhou2022algorithmic,shaki2025bayesian}.
Persuading multiple receivers is a natural generalization of the standard single-receiver persuasion model proposed by \citet{kamenica2011bayesian}.
This generalization introduces many computational challenges. Intractability or inapproximability results have been reported in various related works (e.g., \citep{xu2020tractability,castiglioni2020online}).
These negative results steer studies on optimal multi-receiver persuasion to relaxed objectives, either by considering special settings with additional constraints on the problem instances, or by relaxing the quest for polynomial-time algorithms to less demanding ones.
In the case without externalities among the agents,
\citet{xu2020tractability} explored the special case with a small number of worlds, which we also consider in this paper, and demonstrated the tractability of public persuasion on non-degenerate problem instances. More recently, \citet{castiglioni2023public} proposed bi-criteria approximations of optimal public persuasion in quasi-polynomial-time in the case without externalities, too. Our model is essentially one with externalities since the agents do not make decisions independently of the other agents. Models with externalities strictly generalize those without.
Indeed, one can easily show that our model generalizes standard multi-receiver persuasion models, where agents play actions instead of being matched to a partner.

As for research on stable matching, the stable marriage problem is long known to be tractable via the famous Gale-Shapley algorithm \cite{gale1962college}, the complexity of many variants of this problem remained open until much more recently. The WSM problem, which is closely related to our results, is a typical example. 
When the agents' preferences do not include ties, \citet{vate1989linear} showed the tractability of WSM by formulating it as linear programming. The problem becomes harder when ties or incomplete preference orders may appear. \citet{iwama1999stable} conducted a comprehensive study on stable matching under these circumstances.
It is shown that finding a stable matching is intractable when  both ties and incomplete preference orders may appear, while intractability (more specifically, inapproximability) of the WSM problem only relies on the occurrence of ties. 
We refer the interested readers to an excellent survey of variants of stable matching problems by \citet{iwama2008survey}.

Prior to our work, a few papers had already considered some combination of team formation and Bayesian persuasion. 
The closest to ours is the work by \citet{hssaine2018information}, who studied the problem of persuading one-sided matchings.
Somewhat differently from our model, in theirs, each agent has a type sampled from a prior distribution, which is known to the principal but not to the other agents. Characterizations of optimal policies were presented depending on the convexity/concavity of the utility functions, but no algorithmic results were provided.
It is unknown in what cases efficient algorithms can be designed for computing optimal signaling policies.
Even earlier, \citet{ieong2008bayesian} also examined a coalition formation setting, which is conceptually related to our work, but the model is sufficiently different as they focus on coalition formation in an uncertain world, without any consideration of signaling policy design. Similar works on stable matching under uncertainty were carried out, in which, however, no information signaling is introduced, and the goal is to find a matching that is stable under multiple worlds \citet{aziz2016stable, aziz2022stable, miyazaki2019jointly, genc2017robust, genc2017finding, mai2018finding}. 
\section{Preliminaries}
\label{sc:preliminaries}

In the stable matching problem, a group of $n$ agents $A=\{a_1, \dots, a_n\}$ are to be matched with another group $B=\{b_1, \dots, b_n\}$. Each agent $x \in A \cup B$ gets a value $v_x(y)$ from being matched to another agent $y$ in the other group. 
A matching is given by a bijection $M: A \to B$, where 
With a slight abuse of notation,
we denote by $M(x)$ the matched partner of agent $x$, for $x$ from both $A$ and $B$.
The value $v_x(M)$ of an agent $x$ for a matching $M$ is her value for the agent she is matched to in $M$, i.e., $v_x(M) := v_x(M(x))$.

A matching $M$ is {\em stable} if and only if there does not exist a blocking pair: a pair of agents who prefer each other to their partners in $M$ (\Cref{def:stable-matching}).
For ease of description, we will sometimes also view a matching $M$ as the corresponding set of matched pairs: $\{(a,b) \in A \times B : b = M(a) \}$.

\begin{definition}[Stable matching]
\label{def:stable-matching}
A matching $M$ is {stable} if for every pair of agents $(a,b) \in A \times B$ it holds that
$v_a(M) \ge v_a(b)$ or $v_b(M) \ge v_b(a)$. Otherwise, it is unstable. 
\end{definition}


It is well-known that a stable matching defined above always exists, and there may be multiple. We consider the weighted stable matching setting. In this setting, a principal selects one of the stable matchings for the agents. The principal has her own preference about how the agents should be matched and obtains a utility $u(a, b)$ (i.e., a weight) for each pair $(a,b) \in M$. Overall, the principal's utility for a matching $M$ is the sum $u(M) = \sum_{(a, b) \in M} u(a, b)$.

\subsection{Stable Matching in an Uncertain World}

We consider scenarios with an uncertain world drawn from a distribution $\mu \in \Delta(\Omega)$. The agents' values depend on the realized world. Let $v_x(y \given \omega)$ be the value of agent $x$ for agent $y$ in world $\omega$. Similarly, we augment the principal's utility function and define $u(M, \omega)=\sum_{(a, b) \in M} u(a, b \given \omega)$ as her utility in world $\omega$.
The agents cannot observe $\omega$ directly but know the distribution $\mu$.
On the other hand, the principal knows $\omega$, and the information asymmetry allows her to send signals to reshape the agents' preferences before a matching is selected. The process then proceeds as follows: 
 
\begin{itemize}
\item[1.] Nature draws a world $\omega$ from the distribution $\mu$.

\item[2.] The principal observes $\omega$ and sends a signal $g$ to the agents based on $\omega$; the agents observe $g$ (or part of $g$ in the case of private persuasion).

\item[3.] The principal selects a matching $M$ and the agents are matched accordingly.

\end{itemize}

As in the standard persuasion model, the principal commits to a signaling policy, and the agents update their beliefs based on this commitment as well as information they receive. 
Somewhat differently from the standard persuasion model, in the above process, the information the agents receive consists of both the signal $g$ and the matching $M$.
We refer to $(g, M)$ as a {\em meta-signal}.

\paragraph{Public Persuasion} In the case of public persuasion, the principal sends the same signal to all agents; That is, a policy is a map $\sigma: \Omega \to \Delta(G \times \mathcal{M})$, where $G$ is the set of signals used by the policy and $\mathcal{M} := \{M: A\to B\}$ is the set of all possible matchings. 
More specifically, upon the principal observing the world $\omega$, the policy produces a {meta-signal} $(g, M)$ with probability $\sigma(g,M \given \omega)$, which consists of a signal $g$ to be sent to the agents in Step~2 described above, and a matching $M$ to be selected in Step~3. 
Note that $M$ is also included in the meta-signal because it also reveals information that would influence the agents' beliefs.
We denote by $\sigma(g, M \given \omega)$ the probability that the meta-signal $(g, M)$ is drawn. For notational consistency, we will also denote by $\sigma(\cdot \given \omega)$ the distribution of meta-signals.

From the agents' perspective, knowing $\sigma$ and observing $(g, M)$ leads to the following posterior belief about $\omega$, according to the Bayes' rule:
\begin{align*}
\Pro(\omega \given g, M)
= \frac{\Pro(\omega, g, M)}{\Pro(g, M)} 
= \frac{\mu(\omega) \cdot \sigma(g, M \given \omega)} {\sum_{\omega' \in \Omega} \mu(\omega') \cdot \sigma(g, M \given \omega')},
\end{align*}
where $\Pro$ denotes the probability measure induced by $\sigma$. We denote by $\Pro(\cdot \given g, M)$ the posterior distribution $\left(\Pro(\omega \given g, M) \right)_{\omega \in \Omega}$.

We say that the policy is {\em stable} if it always selects a matching $M$ that is stable with respect to the agents' values under the posterior beliefs, i.e., the expected values $\mathbb{E}_{\omega \sim \Pro(\omega \given g, M)} v_x(y \given \omega)$.
For ease of description, we denote the agents' values under posterior belief $p \in \Delta(\Omega)$ as 
\begin{equation}
\label{eq:post-value}
v_x(y \given p) :=
\mathbb{E}_{\omega \sim p} v_x(y \given \omega).
\end{equation}
We say that a matching is stable with respect to $p$ if it is stable under the values $v_x(y \given p)$.

\begin{definition}[Stable policy]
\label{def:stable-policy}
A policy $\sigma: \Omega \to \Delta(S)$, $S \subseteq G \times \mathcal{M}$, is {\em stable} if for every $\omega \in \Omega$ and every $(g, M) \in S$, the matching $M$ is stable with respect to $\Pro(\cdot \given g, M)$.
\end{definition}

When a stable policy $\sigma$ is applied, the principal's utility is
\[
u(\sigma) := \mathbb{E}_{\omega \sim \mu} \mathbb{E}_{(g, M) \sim \sigma(\cdot \given \omega)} u(M \given \omega).
\]
Our goal is to find the best stable policy that maximizes the expectation among all stable policies.

\paragraph{Private Persuasion} In the case of private persuasion, the principal sends signals through private channels in Step~2, so that every agent receives a private signal not observable to the other agents.
We consider the case where the selection of $M$ in Step~3 is still made public, so the agents know with certainty how they will be matched.

Formally, a policy is now a map $\sigma: \Omega \to \Delta(G^{2n} \times \mathcal{M})$, so that a joint signal $g = (g_x)_{x \in A \cup B} \in G^{2n}$ is drawn in Step~2, where the component $g_x$ is the signal for agent $x$. Receiving $(g_x, M)$ leads to the following posterior belief of agent $x$:
\begin{align}
\label{eq:post}
\hspace{-3mm}
\Pro(\omega \given g_x, M)
= \frac{\mu(\omega) \cdot \sum_{g': g'_x = g_x} \sigma(g', M \given \omega)} {\sum_{\omega' \in \Omega} \mu(\omega') \cdot \sum_{g': g'_x = g_x} \sigma(g', M \given \omega')}.
\end{align}
Using this posterior, the stability notion in \Cref{def:stable-policy} extends naturally to the private case.

We will first focus on public persuasion 
and we refer to a public policy simply as a policy. We will still refer to all private policies explicitly to avoid confusion.

\paragraph{Extension to Many-to-One Matchings} Sometimes, each agent on one side of the matching may be matched to multiple agents on the other side (e.g., for students-to-universities matching). Extending our setting to such cases is trivial; for public persuasion, it simply involves duplicating such agents according to the number of partners they can accept, where each copy ranks others and is, in turn, ranked by them in the same way. For private persuasion, a constraint that all copies of the same agent receive the same information should be added to \Cref{eq:lp-prototype,eq:lp-opt-Vst,eq:LP-S}.

It is more interesting to consider the case where the number of available spots is succinctly represented, hence is exponential in the size of the problem. We analyze this in \Cref{sc:small-types}.

\paragraph{An Example} We present a concrete example to illustrate the above-defined model. Consider the ride-sharing scenario. Each passenger (from set $A$) needs to be matched to a driver (from set $B$) to drive them to their destination. Passengers have preferences over drivers based on factors such as vehicle type, pick-up time, and expected travel duration. Conversely, drivers evaluate passengers based on their willingness to pay and the additional travel time required to accommodate them. Importantly, the state of traffic—which significantly impacts both pick-up and travel times—is unknown to both passengers and drivers at the time of matching; and since it impacts the utilities of the agents, each traffic state corresponds a world $\omega \in \Omega$, distributed according to historical data by $\mu \in \Delta(\Omega)$. The actual state of the world is known only to the principal, who signals the matching recommendation alongside an additional signal, in order to ensure the stability of the matching. Both the matching and the additional signal depend probabilistically on the state of the world.

A numerical illustration can be seen in \Cref{exp:non-optimal}.

\section{Structure of Optimal Policies}
\label{sc:structure-opt-policy}

We first analyze the structure of optimal policies and present several useful results.
In the standard persuasion model it is well known according to the revelation principle that it is without loss of optimality to associate each signal with an action of the agent(s) and view it as an {\em action recommendation}.
It would hence be tempting to think that the same revelation-principle-style result would apply to our matching setting, so that 
it suffices to only signal the matching selected in Step~3 and the signal Step~2 is unnecessary.
We demonstrate that this is {\em not} true:
in the following instance, there is a policy $\sigma: \Omega \to \Delta(G \times \mathcal{M})$ that yields a strictly higher utility than does any policy of the form $\sigma: \Omega \to \Delta(\mathcal{M})$.

\begin{example}
\label{exp:non-optimal}
Let $A = \{a_1, a_2\}$ and $B = \{b_1, b_2\}$.
The are two possible worlds $\omega_1$ and $\omega_2$, with $\mu(\omega_1) = \mu(\omega_2) = 0.5$.
Irrespective of the world, the principal gets utility $1$ for the matched pair $(a_1, b_1)$ and gets $0$ for all other pairs. 
The agents' values are given in the table below.
\begin{table}[h!]
\centering
\small
\begin{tabular}{@{\extracolsep{2pt}}rrrrr}
\toprule
      & \multicolumn{2}{c}{$b_1$} & \multicolumn{2}{c}{$b_2$} \\ \cline{2-3} \cline{4-5} 
      & $\omega_1$  & $\omega_2$  & $\omega_1$  & $\omega_2$  \\ \midrule
$a_1$ & $0$          & $0$           & $-1$          & $2$           \\
$a_2$ & $0$           & $0$           & $1$           & $1$ 
    \\     
\bottomrule
\end{tabular} 
\qquad
\begin{tabular}{@{\extracolsep{2pt}}rrrrr}
\toprule
      & \multicolumn{2}{c}{$a_1$} & \multicolumn{2}{c}{$a_2$} \\ \cline{2-3} \cline{4-5} 
      & $\omega_1$  & $\omega_2$  & $\omega_1$  & $\omega_2$  \\ \midrule
$b_1$ & $1$           & $1$           & $0$         & $0$           \\
$b_2$ & $0$           & $0$           & $-2$           & $1$ 
    \\     
\bottomrule
\end{tabular}
\label{tab:my-table}
\end{table}
\end{example}

Since there are only four agents, there are two possible matchings:
$M_1 =\{(a_1, b_1), (a_2, b_2)\}$ and 
$M_0 =\{(a_1, b_2), (a_2, b_1)\}$, which yield utilities $1$ and $0$, respectively, for the principal.
It can be verified that $M_1$ is stable in both worlds, but not when the worlds are mixed (e.g., with respect to the prior $\mu$). 
Hence, the following deterministic policy is stable:
deterministically, it sends $(\omega_1, M_1)$ in world $\omega_1$, and $(\omega_2, M_1)$ in $\omega_2$.
Namely, the policy always reveals the true world and selects $M_1$.
This yields utility $1$ for the principal.

Now suppose for the sake of contradiction that some policy $\sigma: \Omega \to \Delta(\mathcal{M})$ also achieves utility $1$.
It must be that $\sigma(M_1 \given \omega_1) = \sigma(M_1 \given \omega_2) = 1$ as otherwise the principal would obtain $u(M_0) = 0$ with a positive probability. 
Essentially, since the agents receive signal $M_1$ in both worlds,
the signal is uninformative, upon receiving which the posterior belief the agents derive is the same as the prior: 
$\mu(\omega_1 \given M_1) = \mu(\omega_2 \given M_1) = 0.5$.
Under this belief, $a_1$ and $b_2$ prefer each other to the other agents, so $M_1$ would not be stable and $\sigma$ cannot be a valid policy as a result.

The example also implies that signaling can indeed improve the principal's utility strictly, as compared to the case where a stable matching is chosen under the agents' prior belief.


\subsection{Revelation Principle for Matching Persuasion}

The reason behind the failure of the above policy is that the set of posterior distributions that make a matching stable is non-convex.
E.g., in \Cref{exp:non-optimal}, the set that makes $M_1$ stable contains $p = (1,0)$ and $p' = (0,1)$ but not $(p+p')/2$. 
Hence, we cannot merge signals inducing the same stable matching to show that it suffices to use only one signal for every matching.
To obtain a revelation-principle style result for our setting therefore requires a finer division of the posterior space. 
It turns out that this can be done by incorporating the agents' preferences in the signals.

In what follows, we denote by $\preceq := (\preceq_{x})_{x\in A\cup B}$ a preference profile of the agents, where each $\preceq_x$ is a total order on $A$ or $B$ depending on the side of $x$.
Let $\mathcal{L}$ be the set of all possible preference profiles.
A partition of the posterior space $\Delta(\Omega)$ into {\em cells} can be defined based on preference profiles induced by posterior distributions.

\begin{definition}[Cell]
\label{def:cell}
The cell $C_\preceq$ of a preference profile $\preceq$ is the set of 
vectors $p \in \mathbb{R}^{|\Omega|}$ inducing $\preceq$, i.e., $p \in C_\preceq$ if 
\begin{align}
\label{eq:C-prec-p}
v_x(y \given p) \le v_x(y' \given p), \quad \text{for all } x,y,y': y \preceq_x y'.
\end{align}
\end{definition}

A useful observation, by \Cref{eq:C-prec-p} and \Cref{def:stable-matching}, is that posteriors in a cell induce the same set of stable matchings.

\begin{observation}
A matching $M$ is stable with respect to $p$ if and only if it is stable with respect to some $\preceq$ such that $p \in C_\preceq$, i.e., $b \preceq_a M(a)$ or $a \preceq_b M(b)$ for every $(a,b) \in A \times B$.
\end{observation}



Intuitively, since every cell is convex by definition, we can merge meta-signals containing the same matching $M$ and inducing posteriors in the same cell. The merged meta-signal falls in the same cell, so as long as the original meta-signals are {\em indicative} of the agents' preferences (\Cref{def:indicative-policy}) and make $M$ stable, the same holds for the merged one. 

We say that a policy is {\em indicative} if the signals it sends are indicative of the preference profiles induced:

\begin{definition}[Indicative policy]
\label{def:indicative-policy}
A policy $\sigma: \Omega \to \Delta(S)$, $S \subseteq \mathcal{L} \times \mathcal{M}$, is {\em indicative} if $\Pro(\cdot \given \preceq, M) \in C_\preceq$ for every $(\preceq,M) \in S$. 
\end{definition}

Intuitively, an indicative policy is analogous to a direct signaling strategy as in the standard persuasion model (where signals directly indicate actions for the agents to take), whereas stability of a policy is analogous to incentive compatibility. This leads to the following result, which resembles the revelation principle for the standard persuasion model.

\begin{restatable}{theorem}{thmrpC}
\label{thm:rp-C}
There exists an optimal policy ${\sigma}: \Omega \to \Delta(\mathcal{L} \times \mathcal{M})$.
Moreover, ${\sigma}$ is indicative.
\end{restatable}

\begin{proof}
Consider an arbitrary optimal policy $\sigma^*: \Omega \to \Delta(G \times \mathcal{M})$. We will construct an indicative policy ${\sigma}: \Omega \to \Delta(\mathcal{L} \times \mathcal{M})$ that achieves as much utility for the principal as $\sigma^*$ does.
To ease the exposition, we assume that $G$ is a finite set. The result can be easily extended to the case with an infinite $G$ under mild integrability assumptions.

To construct ${\sigma}$, we let $f: G \times \mathcal{M} \to \mathcal{L}$ be a (arbitrary) function that maps every meta-signal $(g, M)$ of $\sigma$ to a preference profile $\preceq$ such that
$\Pro^*(\cdot \given g, M) \in C_\preceq$, where $\Pro^*$ denotes the probability measure induced by $\sigma^*$.
We then merge signals corresponding to the same preference profiles, so that whenever $\sigma^*$ sends $(g,M)$, ${\sigma}$ sends $(f(g, M), M)$. 
Since the merging procedure does not change the matching specified by the meta-signal, the new policy ${\sigma}$ has the same marginal probability for every matching $M$.
This means that as long as ${\sigma}$ is stable, it will yield the same utility for the principal. We show that ${\sigma}$ is stable and indicative next.

The stability of ${\sigma}$ relies on the convexity of the cells, as well as the linearity of the value functions. 
Specifically, let ${\Pro}$ denote the probability measure induced by ${\sigma}$.
We have 
\begin{align*}
{\Pro}(\omega \given \preceq, M)   
= \frac{{\Pro}(\preceq, M, \omega)} {{\Pro} (\preceq, M)} 
= \frac{\sum_{(g, M): f(g, M) = \preceq}
\Pro (g, M, \omega)} {{\Pro} (\preceq, M)},
\end{align*}
where we used the fact that ${\Pro}(\preceq, M, \omega) = \sum_{(g,M):f(g,M) = \preceq} \Pro (g, M, \omega)$ due to the way the meta-signals are merged.
The right-hand side can further be written as:
\begin{align*}
\sum_{(g, M): f(g, M) = \preceq} 
\frac{\Pro^* (g, M)}{{\Pro} (\preceq, M)} 
\cdot 
\frac{\Pro^* (g, M, \omega)} {\Pro^* (g, M)} 
= \sum_{(g, M): f(g, M) = \preceq} 
\frac{\Pro^* (g, M)}{{\Pro} (\preceq, M)} 
\cdot 
\Pro(\omega \given g, M).
\end{align*}
By construction we have $\sum_{(g, M): f(g, M) = \preceq} 
\frac{\Pro^* (g, M)}{{\Pro} (\preceq, M)} = 1$, so the above means that the posterior distribution ${\Pro}(\cdot \given \preceq, M)$ is a convex combination of the set of vectors $\{ \Pro^* (\cdot \given g, M) : f(g, M) = \preceq \}$. 

By definition, $f(g, M) = \preceq$ implies that $\Pro^*(\cdot \given g, M) \in C_\preceq$, so all these vectors are in $C_\preceq$ and so is ${\Pro}(\cdot \given \preceq, M)$.
Hence, ${\sigma}$ is indicative.
Since $\sigma^*$ as an optimal policy must be stable, $M$ is stable with respect to $\preceq$.
As a result, ${\sigma}$ is also stable. This completes the proof.
\end{proof}

\subsection{Computing Optimal Policies}

Based on \Cref{thm:rp-C}, a linear program (LP) can be formulated as follows to compute optimal policies. 
We consider the set of stable policies of the form $\sigma: \Omega \to \Delta(\mathcal{L} \times \mathcal{M})$.
Such policies can be represented by the set of variables $V = \{\sigma(\preceq, M \given \omega) : \preceq \in \mathcal{L}, M \in \mathcal{M}, \omega \in \Omega \}$. 
\begin{align}
\label{eq:lp-opt}
\max\ 
\sum_{\omega \in \Omega} \mu(\omega) \sum_{(\preceq,M) \in \mathcal{L} \times \mathcal{M}} \sigma(\preceq, M \given \omega) \cdot u(M \given \omega),
\end{align}
subject to the following (in addition to $\sigma(\cdot \given \omega) \in \Delta(\mathcal{L} \times \mathcal{M})$):
\begin{itemize}
\item 
For each $(\preceq, M) \in \mathcal{L} \times \mathcal{M}$, we require
$\Pro(\cdot \given \preceq,M) \in C_\preceq$, to ensure that the resulting policy is indicative and allow filtering for stability.
According to \Cref{eq:C-prec-p,eq:post,eq:post-value}, this is equivalent to the following linear constraint for all $(\preceq, M) \in \mathcal{L} \times \mathcal{M}$ and all $x,y,y'$ such that $y \preceq_x y'$:
\begin{align}
& \sum\nolimits_{\omega} \mu(\omega) \cdot \sigma(\preceq,M \given \omega) \cdot v_x(y \given \omega) 
\le \sum\nolimits_{\omega} \mu(\omega) \cdot \sigma(\preceq,M \given \omega) \cdot v_x(y' \given \omega) 
\tag{\ref{eq:lp-opt}-1}
\label{eq:lp-opt-1}
\end{align}

\item 
Finally, for every $\preceq \in \mathcal{L}$ and $M \in \mathcal{M}$ unstable with respect to $\preceq$:
\begin{equation}
\label{eq:lp-opt-2}
\sigma(\preceq, M \given \omega) = 0.
\tag{\ref{eq:lp-opt}-2}
\end{equation}

\end{itemize}
\noindent
Jointly, \Cref{eq:lp-opt-1,eq:lp-opt-2} ensure that $M$ is stable with respect to $\Pro(\cdot \given \preceq,M)$ if ever $(\preceq, M)$ is sent with a nonzero probability.
Namely, if $\sigma(C_\preceq, M \given \omega) > 0$, then \Cref{eq:lp-opt-2} implies that 
$M(a) {\preceq}_a b$ and $M(b) {\preceq}_b a$ for all $(a, b)$,
in which case
\Cref{eq:lp-opt-1} further implies the stability of $M$ according to \Cref{def:stable-matching}.
The correctness of the LP formulation then follows readily by \Cref{thm:rp-C}.

Unfortunately, the above LP does not lead to any efficient algorithm as the sets $\mathcal{L}$ and $\mathcal{M}$ grow exponentially with $n$. 
Indeed, even when there is no uncertainty about the world, the problem is hard due to the intractability of the WSM (weighted stable matching) problem (in the case where ties are allowed in the agents' preference orders) \cite{iwama1999stable}. 
Given this inapproximability result, we investigate special cases of the problem in the next sections.

\begin{theorem}[{\citealp[Theorem~2 rephrased]{iwama1999stable}}]
\label{thm:nph}
Assuming P$\neq$NP, there exists no efficient algorithm that computes a $\frac{1}{n^{1- \epsilon}}$-approximation of an optimal public policy for any constant $\epsilon > 0$, even when $|\Omega| = 1$.
\end{theorem}

Hence, in the next sections we consider the computational complexity of various relaxed settings.

\section{Small Number of Agent Types}
\label{sc:small-types}

Sometimes the agents' preferences only have a small number of types. 
Agents of the same type value other agents the same way, while they are also valued the same by other agents. Somewhat similar models where considered in \cite{shrot2010agent, shaki2025bayesian}.

For example, in the student-to-university setting, students can be categorized according to their grades (e.g., A, B, C, D) and interests. Each university is modeled as the multiple available seats it has, each with the same preferences as the university, and with all seats in the same university being identical for the students. Hence, the available seats are already categorized into a small number of options, by the university they belong to. 

\subsection{A Type-Based Model}

Formally, let there be $T$ types of agents on each side.
Let $\mathcal{A} = \{ A_1, \dots, A_T\}$ and $\mathcal{B} = \{ B_1, \dots, B_T\}$ be partitions of $A$ and $B$ into $T$ subsets, so that each subset consists agents of one type.
The value functions are now defined based on the agents' types.
For any agents $x$ and $y$ of types $s$ and $t$, respectively, we have
$v_x(y \given \omega) = v_s(t \given \omega)$
for all $\omega$.
Similarly, when $x$ and $y$ are matched together, the principal's utility for the pair $(x,y)$ depends only on $s$ and $t$, instead of the agents' specific identities:
$u(x,y \given \omega) = u(s, t \given \omega)$ 
for all $\omega$.
Only the values $v_s(t \given \omega)$, $u(s, t \given \omega)$ are given in the problem instance.

\paragraph{Exponentially Many Agents}
The type-based model admits a succinct representation that allows the number of agents to grow exponentially in the size of the problem instance. To define the agents' values, it suffices to specify, for each agent type, the value function of this type and its size.
The sizes can be encoded in binary and hence grow exponentially in the bit-size of the representation. We denote by $\eta(t)$ the size of type $t \in \mathcal{A} \cup \mathcal{B}$. 
As we demonstrate next, our algorithm is capable of handling this succinct representation; its run time grows only polynomially in $\log n$. 

\subsection{Prototypes and Useful Matchings}

We introduce {\em prototypes} and define useful matchings based on them, which will serve as a key component of our algorithm.
A prototype is given by a set $P \subseteq \mathcal{A} \times \mathcal{B}$ of type pairs. It represents the class of matchings where the occurrence of type pairs is exactly the same as the pairs in $P$: no pairs are matched if their types are not in $P$.

\begin{definition}[Prototype]
\label{def:prototype}
The prototype of a matching $M$ is the set $\{(A_i, B_j) \in \mathcal{A} \times \mathcal{B}: M \cap (A_i \times B_j) \neq \varnothing \}$.
\end{definition}

For any $P \subseteq \mathcal{A} \times \mathcal{B}$, we let $\mathcal{M}_P$ denote the class of matchings whose prototype is $P$. 
Prototypes are useful because matchings in each $\mathcal{M}_P$ are either all stable or all unstable, with respect to some posterior distribution. 
More specifically, we make the following observation.

\begin{observation}
\label{lmm:stability-prototype}
Suppose $M \in \mathcal{M}_P$, $M' \in \mathcal{M}_{P'}$, and $P' \subseteq P$.
If $M$ is stable with respect to posterior $p \in \Delta(\Omega)$, then $M'$ is also stable with respect to $p$.
\end{observation}

Hence, given a policy $\sigma$, we can substitute the matching $M$ selected by a meta-signal $(\preceq, M)$ with any $M' \in \mathcal{M}_{P'}$, without changing the stability of $\sigma$.
We will use this property to 
reduce the size of the necessary signal space by substituting matchings with a small set of useful matchings.
Specifically, the best matching to replace $M$ can be characterized by the following optimization:
\begin{equation}
\label{eq:opt-M-prototype}
\max_{M' \in \mathcal{M}_{P'},\ P' \subseteq P}\quad
\sum_{\omega \in \Omega} \mu(\omega) \cdot \sigma(\preceq, M \given \omega) \cdot u(M' \given \omega),
\end{equation}
where the objective corresponds to the part in \Cref{eq:lp-opt} that depends on the meta-signal $(\preceq, M)$. 
This optimization can further be formulated as the following LP, where $\{M'(s,t): s \in \mathcal{A}, t \in \mathcal{B}\}$ is the set of variables, and we write $q(\omega) := \mu(\omega) \cdot \sigma(\preceq, M \given \omega)$.

\begin{align}
\label{eq:lp-prototype}
\max\quad 
\sum_{\omega \in \Omega} 
q(\omega)
\sum_{(s, t) \in \mathcal{A} \times \mathcal{B}} M'(s,t) \cdot u(s, t \given \omega) 
\end{align}
subject to:
\begin{align}
&\textstyle 
\sum_{t \in \mathcal{B}} M'(s, t) = \eta(s) && \text{for all } s \in \mathcal{A} \tag{\ref{eq:lp-prototype}-1} \label{eq:lp-prototype-1}\\
&\textstyle 
\sum_{s \in \mathcal{A}} M'(s, t) = \eta(t) && \text{for all } t \in \mathcal{B} \tag{\ref{eq:lp-prototype}-2} \label{eq:lp-prototype-2}\\
& M'(s, t) \ge 0 && \text{for all } (s, t) \in P \tag{\ref{eq:lp-prototype}-3} \label{eq:lp-prototype-3}\\
& M'(s, t) = 0 && \text{for all } (s, t) \notin P \tag{\ref{eq:lp-prototype}-4} \label{eq:lp-prototype-4}
\end{align}
Specifically, every feasible solution $M'$ according to the above formulation corresponds to a class of matchings in which the number of type-$(s,t)$ pairs is exactly $M'(s,t)$ for every $(s,t) \in \mathcal{A} \times \mathcal{B}$. 
Due to the anonymity in the principal's utility function, all these matchings yield the same utility for the principal, so the specific identities of the agents are irrelevant.
\Cref{eq:lp-prototype-1,eq:lp-prototype-2,eq:lp-prototype-3} ensure that the numbers of agents specified by $M'$ are consistent with the size of each type.
\Cref{eq:lp-prototype-4} ensures that the prototype of $M'$ is a subset of $P$.

Note that the agent numbers specified by $M'$ must be integers, but there is no need to force variables in the LP to be integers because there always exists an integer optimal solution according to the following lemma. The lemma is due to \Cref{eq:lp-prototype-1,eq:lp-prototype-2,eq:lp-prototype-3,eq:lp-prototype-4} being network flow constraints. The integrality theorem 
states that a polytope defined by such constraints has only integer vertices.

\begin{restatable}{lemma}{lmmIntegersVertices}
\label{lmm:intergers-vertices}
The vertices of the feasible set of LP~\eqref{eq:lp-prototype} are all integer solutions. 
\end{restatable}

\begin{proof}
Note that \Cref{eq:lp-prototype-1,eq:lp-prototype-2,eq:lp-prototype-3,eq:lp-prototype-4} are network flow constraints: we can view $M(s,t)$ as the amount of flow between $s$ and $t$; and view $\eta(s)$ and $\eta(t)$ as capacities of the edges from the source to $s$ and the edges from $t$ to the sink, respectively. So, according to the integrality theorem (see, e.g., \citep[Theorem~13.2]{vanderbei2020linear}), 
the vertices of the polytope are all integer vectors.
Since every LP has at least one optimal solution that is a vertex of the feasible region, the claimed result then follows.
\end{proof}

\subsection{Efficient Algorithm for Small \texorpdfstring{$T$}{T}}


We can formulate LP~\eqref{eq:lp-prototype} for every meta-signal $(\preceq, M) \in \mathcal{L} \times \mathcal{M}$ to find out a substitute of $M$.
A key property of the formulation is that the feasible region depends only on the prototype of $M$:

\begin{observation}
The feasible region of LP~\eqref{eq:lp-prototype} depends only on $P$.
It is independent of $\sigma$.
\end{observation}

\noindent
Hence, we let $\mathcal{V}_P$ denote the {\em vertex set} of LP~\eqref{eq:lp-prototype}.
Since every LP has at least one optimal solution that is a vertex of the feasible region, $\mathcal{V}_P$ then serves as a candidate set universal for all $M \in \mathcal{M}_P$: it contains at least one best substitute for every $M \in \mathcal{M}_P$.
Enumerating all possible prototypes, we further obtain the set $\mathcal{V}^* = \bigcup_{P \in \mathcal{A}\times\mathcal{B}} \mathcal{V}_P$ that contains all of the useful matchings we need for designing an optimal policy. This then gives the following result with a tighter bound on the necessary signal space for the type-based model.
Indeed, the size of $\mathcal{V}^*$ is constant given a constant number of types (as the size of LP~\eqref{eq:lp-prototype} is constant), and this is the cornerstone of the efficient algorithm we present next.

\begin{restatable}{theorem}{thmRpPrototype}
\label{thm:rp-prototype}
There exists an optimal policy ${\sigma}: \Omega \to \Delta(\mathcal{L} \times \mathcal{V}^*)$ that is also indicative.
\end{restatable}

\begin{proof}
The proof is similar to that of \Cref{thm:rp-C}.
Without loss of generality, we can assume that there is an optimal policy $\sigma^*: \Omega \to \Delta(\mathcal{L} \times \mathcal{M})$ according to \Cref{thm:rp-C}.
To construct ${\sigma}$, we replace each meta-signal $(\preceq, M)$ of $\sigma$ by a meta-signal $(\preceq, M')$ such that $M'$ is an (arbitrary) optimal solution of LP~\eqref{eq:lp-prototype}.
This yields a policy ${\sigma}: \Omega \to \Delta(\mathcal{L} \times \mathcal{V}^*)$. It can further be verified that ${\sigma}$ is stable and indicative via similar arguments in the proof of \Cref{thm:rp-C}.
\end{proof}



Given \Cref{thm:rp-prototype}, we then 
use the same formulation as LP~\eqref{eq:lp-opt} to compute an optimal policy, where we replace the meta-signal space $\mathcal{L} \times \mathcal{M}$ to the smaller set $\mathcal{L} \times \mathcal{V}^*$.
The LP can be solved at the type level as constraints involving agents of the same types are essentially the same. This amounts to the following version of LP~\eqref{eq:lp-opt} where we let $\mathcal{L} = (L_{\mathcal{B}})^T \times (L_{\mathcal{A}})^T$, with $L_{\mathcal{A}}$ and  $L_{\mathcal{B}}$ being the sets of total orders on $\mathcal{A}$ and $\mathcal{B}$, respectively: 
\begin{align}
\label{eq:lp-opt-Vst}
&\max\ 
\sum_{\omega \in \Omega} \mu(\omega) \sum_{(\preceq,M) \in \mathcal{L} \times \mathcal{V}^*} \sigma(\preceq, M \given \omega) \cdot u(M \given \omega),
\end{align}
subject to the following constraints for all $(\preceq, M) \in \mathcal{L} \times \mathcal{V}^*$:

\begin{itemize}
\item 
$\sum\nolimits_{\omega} \mu(\omega) \cdot \sigma(\preceq,M \given \omega) \cdot \Big( v_x(y \given \omega) -  v_x(y' \given \omega) \Big) \le 0$, for all 
$x,y,y' \in \mathcal{A} \cup \mathcal{B}: y \preceq_x y'$.

\item
$\sigma(\preceq, M \given \omega) = 0$,
for all $\omega \in \Omega$ and $M \in \mathcal{V}^*$ unstable w.r.t. $\preceq$.

\item 
$\sigma(\cdot \given \omega) \in \Delta(\mathcal{L} \times \mathcal{V}^*)$
for all $\omega \in \Omega$.
\end{itemize}
Due to the small size of $\mathcal{V}^*$
when $T$ is small, LP~\eqref{eq:lp-opt-Vst} can be solved efficiently. This yields an efficient algorithm, which we summarize in \Cref{alg:small-T}.
The algorithm runs in $\poly(\log n)$ time, so it is efficient even under the succinct representation of the type-based model with exponentially many agents.

\begin{algorithm}[t]
\caption{Compute an optimal policy (when $T$ is small)\label{alg:small-T}}

\begin{enumerate}[leftmargin=3mm]
\item 
Compute $\mathcal{V}_P$ for each prototype $P \subseteq \mathcal{A} \times \mathcal{B}$. 

\item 
Solve LP~\eqref{eq:lp-opt-Vst}, where $\mathcal{V}^* = \bigcup_{P \in \mathcal{A}\times\mathcal{B}} \mathcal{V}_P$.
\end{enumerate}
\end{algorithm}

\begin{restatable}{theorem}{thmTypeModelTimeComplexity}
\label{thm:type-model-time-complexity}
An optimal policy can be computed in $\poly(|\Omega|, \log n)$ time when $T$ is a constant.
\end{restatable}

\begin{proof}
Given \Cref{thm:rp-prototype}, we use LP~\eqref{eq:lp-opt-Vst} to compute an optimal policy.

The LP has $|\Omega| \cdot |\mathcal{L}| \cdot |\mathcal{V}^*|$ variables and $O(|\Omega| \cdot |\mathcal{L}| \cdot |\mathcal{V}^*| \cdot T^3)$ constraints, so it can be solved in time polynomial in these parameters.

When $T$ is a constant, $|\mathcal{L}|$ is also a constant as $\mathcal{L}$ is now defined at a type level.
We analyze $|\mathcal{V}^*|$ next.

Consider LP~\eqref{eq:lp-prototype}, which defines $\mathcal{V}_P$. The size of LP~\eqref{eq:lp-prototype} is constant when $T$ is a constant. So, each $\mathcal{V}_P$ only contains a constant number of vertices. There are only a constant number of prototypes, too, so $|\mathcal{V}^*|$ is constant.
To obtain each $\mathcal{V}_P$, we enumerate every possible combination of hyperplanes that define the feasible region, and compute their intersection by solving the corresponding system of equations. 
This takes time $\poly(\log n)$ for each prototype $P$, where the $\log n$ term arises because of the parameters $\eta(s)$ and $\eta(t)$ in the equations.
Since there is a constant number of prototypes when $T$ is constant, the total amount of time it takes to compute $\mathcal{V}^*$ is also $\poly(\log n)$.

In summary, the time complexity of the algorithm is $\poly(|\Omega|, \log n)$ when $T$ is a constant.
\end{proof}

\section{Small Number of Worlds}
\label{sc:small-worlds}

We now turn back to the model without types and the case with a small number of worlds.
Recall that because of ties in the agents' preferences, optimal policies are hard to compute even when there is only one world.
In the persuasion model, ties are even more ubiquitous: even when an agent has strict preferences in every world, ties may arise when she only holds a probabilistic belief about which world the current one is. As a result, a non-degeneracy condition is necessary for our efficient algorithm. We define this condition first.

\subsection{Non-degeneracy}

In what follows, we denote by $\prec = (\prec_x)_{x \in A \cup B}$ a {\em strict} preference profile, which consists of a strict order $\prec_x$ for every agent $x$. 
We define the following non-degenerate condition and will argue in \Cref{thm:adding-noise} that the definition is well-motivated in the sense that non-degenerate instances emerge with probability $1$ if parameters of an instance are drawn at random, or if a small random noise is imposed on each parameter. 

\begin{definition}[Non-degeneracy]
\label{def:non-degeneracy}
Given a strict preference profile $\prec$, let $V_\prec = \bigcup_{x \in A \cup B} V_\prec^x$, where 
\[
V_\prec^x := 
\Big\{ \big(v_x(y_\ell \given \omega) - v_x(y_{\ell+1} \given \omega) \big)_{\omega \in \Omega}\ :\ \ell = 1,\dots, n-1
\Big\}
\]
with $y_1 \prec_x y_2 \prec \dots \prec_x y_n$.
A problem instance is {\em non-degenerate} if, for any order $\prec$, the vectors in any size-$|\Omega|$ subset of $V_\prec$ are linearly independent.
\end{definition}

We remark that a similar non-degeneracy condition was introduced by \citet{xu2020tractability} for a standard public persuasion model, but with a fundamental difference that it is not defined with respect to any orders. 
Indeed, orders are not necessary in their case because the number of the agent's actions is restricted to two in their model. 
They become necessary when there are three or more actions (even in the model of \citet{xu2020tractability}). Specifically, without the orders in the definition, one would include all vectors $\mathbf{v}_{ij}^x := (v_x(y_i \given \omega) - v_x(y_j \given \omega))_\omega$ in $V^x$ for all distinct $i,j$, in which case any three vectors $\mathbf{v}_{ij}^x, \mathbf{v}_{jk}^x, \mathbf{v}_{ki}^x$ would be linearly dependent as $\mathbf{v}_{ij}^x + \mathbf{v}_{jk}^x + \mathbf{v}_{ki}^x \equiv \mathbf{0}$;\footnote{This does not happen in a public persuasion model with only two actions, say actions $1$ and $2$, because only one vector $\mathbf{v}_{12}^x$ is included in the set for every agent $x$.} 
consequently, all instances would be considered {\em degenerate} and degenerate instances would not constitute a zero measure anymore. 
Additional arguments are therefore also necessary for the proofs of \Cref{thm:adding-noise} and \Cref{lmm:Cpp}. These treatments might be of independent interest to research on generic public persuasion models with arbitrary number of actions.

\begin{restatable}{proposition}{thmAddingNoise}
\label{thm:adding-noise}
Let $\tilde{v}_x(y|\omega) = v_x(y|\omega) + \epsilon_{x, y, \omega}$ where $\epsilon_{x, y, \omega}$ is a random variable following the uniform distribution $U(-\epsilon, \epsilon)$. For any $\epsilon > 0$ and any value function $v$, instances with value function $\tilde{v}$ is non-degenerate with probability 1.
\end{restatable}

\begin{proof}
Suppose that $\tilde{v}$ leads to a degenerate instance. 
By definition, there exist a strict preference profile $\prec$ and a set $\widetilde{V} = \bigcup_{x\in A\cup B} \widetilde{V}_x$ consisting of 
$|\Omega|$ linearly dependent vectors, where each $\widetilde{V}_x$ contains vectors in the form
$(\tilde{v}_{x}(y_{\ell} \given \omega) - \tilde{v}_{x}(y_{\ell+1} \given \omega))_{\omega \in \Omega}$ such that $y_{1} \prec_{x} y_{2} \prec_{x} \dots \prec_{x} y_{n}$.
Viewing $\widetilde{V}$ as a matrix whose columns are the vectors contained in $\widetilde{V}$ as a set, we have 
\[
\widetilde{V} \cdot \mathbf{a} = \mathbf{0}
\]
for some {\em nonzero} column vector $\mathbf{a}$, whose components are not all zero. Put it differently, we have
\[
\widetilde{V}_x \cdot \mathbf{a}_x = \mathbf{0}
\]
for every $x$, where we view $\widetilde{V}_x$ as a matrix the same way we view $\widetilde{V}$ and $\mathbf{a}_x$ is a vector consisting of components in $\mathbf{a}$ that corresponds to the rows of $\widetilde{V}_x$ as in $\widetilde{V}$. 

Since $\mathbf{a}$ is nonzero, at least one $\mathbf{a}_x$ is also nonzero. 
Let us pick such an $x$. 
Note that we can further group the rows of $\widetilde{V}_x$, so that we divide it into sub-matrices of the form
\[
(\tilde{\mathbf{v}}_j - \tilde{\mathbf{v}}_{j+1},\
\tilde{\mathbf{v}}_{j+1} - \tilde{\mathbf{v}}_{j+2},\
\dots,\
\tilde{\mathbf{v}}_{j+k-1} - \tilde{\mathbf{v}}_{j+k})^{\mathsf{T}},
\]
where $\tilde{\mathbf{v}}_j := (\tilde{v}_x(y_j \given \omega))_{\omega \in \Omega}$.
Moreover, at least one of these sub-matrices must correspond to a nonzero sub-vector of $\mathbf{a}_x$.
Let this sub-vector be $(b_1, \dots, b_k)$. Then we get that 
\begin{equation}
\label{eq:bv-linear-dependent}
b_1 \tilde{\mathbf{v}}_j +
(b_2 - b_1) \tilde{\mathbf{v}}_{j+1} + \dots
(b_k - b_{k-1}) \tilde{\mathbf{v}}_{j+k-1}
- b_k \tilde{\mathbf{v}}_{j+k} 
= \mathbf{0}.
\end{equation}
The coefficients above cannot all be zero because when they are all zero, $(b_1, \dots, b_k)$ must be a zero vector as well. 
Hence, $\tilde{\mathbf{v}}_j,\dots, \tilde{\mathbf{v}}_{j+k}$ are linearly dependent.

Note that $k + 1 \le |\Omega|$ as there are only $|\Omega|$ vectors in $\widetilde{V}$.
It can then be shown via induction that the probability of drawing $\tilde{\mathbf{v}}_j,\dots, \tilde{\mathbf{v}}_{j+k}$ that are linearly dependent is $0$.
Specifically, as the base case, the probability of getting $\tilde{\mathbf{v}}_j = \mathbf{0}$ is zero. Now suppose the probability of the first $\ell$ vectors $\tilde{\mathbf{v}}_j,\dots, \tilde{\mathbf{v}}_{j+\ell-1}$ being linearly dependent is $0$.
For $\tilde{\mathbf{v}}_j,\dots, \tilde{\mathbf{v}}_{j+\ell}$ to be linearly independent, $\tilde{\mathbf{v}}_{j+\ell}$ must be a linear combination of the first $\ell - 1$ vectors. The probability of this happening is zero as well because the measure of the span of the first $\ell - 1$ vectors is zero. This then implies the desired result. 
\end{proof}

\subsection{Using Proper Cells}
\label{sc:proper-cells}

The non-degeneracy assumption allows us to consider only {\em proper cells} (as an extension of cells in \Cref{def:cell}) to restrict the space of preference profiles to be considered. This is essential as the number of proper cells grows only polynomially with $n$ when $\Omega$ is small.
Additionally, as we will see later, the associated {\em strict} preferences of the proper cells also allow us to utilize a tractability result of the WSM (weighted stable matching) problem in the case without ties in the agent's preferences.


\begin{definition}[Proper cell]
\label{def:proper-cell}
The cell $C_\prec$ of a {\em strict} preference profile $\prec$, called a {\em proper cell}, is the closure of the set of vectors
$p \in \mathbb{R}^{|\Omega|}$ that induces $\prec$,
i.e., $p \in C_\prec$ if and only if $v_x(y \given p) < v_x(y' \given p)$ for all $x,y,y': y \prec_x y'$.
\end{definition}

\noindent
It is straightforward according to the definition that $p \in C_\prec$ if and only if
\begin{align}
\label{eq:C-prec-p-2}
v_x(y \given p) \le v_x(y' \given p), \quad \text{for all } x,y,y': y \prec_x y'.
\end{align}
Consider the collection of all {\em nonempty} proper cells that arise in the problem instance, and we denote by $\mathcal{L}^\proper$ the set of strict preference profiles corresponding to them. 
By definition, a proper cell $C_\prec$ is the same as $C_\preceq$, where $\preceq$ is the reflexive closure of $\prec$ given by: $y \preceq_x y'$ if $y \prec_x y'$ or $y = y'$.
Moreover, proper cells are exactly the cells whose interior is nonempty.
As pointed out by \citet{xu2020tractability}, the following upper bound on the number of proper cells can be established via a classic result on how many cells can a given number of hyperplanes divide an euclidean space of a fixed dimension into.

\begin{restatable}{lemma}{lmmNumProperCells}
\label{lmm:num-proper-cells}
$|\mathcal{L}^\proper| = O(n^{3|\Omega|})$.
Moreover, $\mathcal{L}^\proper$ can be calculated efficiently when $|\Omega|$ is small. 
\end{restatable}

\begin{proof}
As pointed out by \citet{xu2020tractability},
$h$ hyperplanes divide $\mathbb{R}^d$ into at most $O(h^d)$ cells with nonempty interiors according to a classic result in mathematics \cite{steiner1826einige}. In our case, there are at most $n^3$ hyperplanes (see \Cref{eq:C-prec-p}) dividing the space $\mathbb{R}^{|\Omega|}$. The upper bound then follows. 

The cells can further be identified via the following algorithm (see also \cite{xu2020tractability}).
Specifically, the algorithm adds hyperplanes one by one. A collection of cells is maintained to store cells that have been generated so far. Each time a new hyperplane is added, the algorithm enumerates each cell in the collection to check if it is divided by the new hyperplane into two, or it remains the same. In the former case, the newly generated cells are added into the collection to replace the old ones. Because of the upper bound on the number of cells that can be generated via this process, the algorithm runs efficiently when $|\Omega|$ is small.
\end{proof}

Additionally, the following lemma indicates that it is without loss of generality to consider only proper cells and strict orders when we compute optimal policies. This is where the non-degeneracy comes into play.
Hence, together with \Cref{lmm:num-proper-cells}, this reduces the number of cells we need to consider to a polynomial number in the case of small $|\Omega|$.


\begin{restatable}{lemma}{llmCpp}
\label{lmm:Cpp}
Given a non-degenerate problem instance, there exists an optimal policy $\sigma$ such that $\sigma: \Omega \to \Delta(\mathcal{L}^\proper \times \mathcal{M})$ and $\sigma$ is indicative. 
\end{restatable}

\begin{proof}
Suppose that $\sigma^*: \Omega \to \Delta(S^*)$ is an optimal policy supported on some $S^*$. By \Cref{thm:rp-C}, we can assume that $S^* \subseteq \mathcal{L} \times \mathcal{M}$ and $\sigma^*$ is indicative.

We replace $\preceq$ in each $s = (\preceq,M) \in S^*$ by some $\prec \in \mathcal{L}^\proper$ to construct $\sigma$.
Specifically, 
consider the posterior $p := \Pro^*(\cdot \given s)$ induced by $s$ (under $\sigma^*$).
We will replace $\preceq$ by some $\prec$ such that $p \in C_\prec$.
Indeed, according to the definition of proper cells, the collection of all proper cells forms a partition of $\Delta(\Omega)$, 
so there must be at least one such $\prec$.
Via the same argument in the proof of \Cref{thm:rp-C}, it can also be seen that the policy remains stable and indicative after the replacement, for any choice of $\prec$ such that $p \in C_\prec$.
It remains to show that there exists a choice of $\prec$ such that, in addition to $p \in C_\prec$, $M$ is also stable with respect to $\prec$.

\paragraph{Claim (1).} {\em 
$M$ is stable with respect to some $\prec \in \mathcal{L}^+$ consistent with $\preceq$, i.e., some $\prec$ such that $y \prec_x y'$ only if $y \preceq_x y'$.}

\smallskip

Indeed, since $\sigma^*$ is stable, $M$ must be stable with respect to $p$. 
A strict profile $\prec$ that makes $M$ stable can be constructed as follows. First, for every tuple $(x, y, y')$, we let $y \prec_x y'$ if $v_x(y\given p) < v_x(y' \given p)$, or $v_x(y\given p) = v_x(y' \given p)$ and $y' = M(x)$. 
Then, set all other unspecified relations arbitrarily.
Note that $p \in C_\preceq$ as $\sigma^*$ is indicative. Hence, the above constructed $\prec$ is consistent with $\preceq$.

\smallskip

Given Claim~(1), we next demonstrate that $p \in C_\prec$ for every $\prec$ consistent with $\preceq$ to complete the proof.
Intuitively, every $\prec$ results from a feasible way of breaking ties in $\preceq$.
Specifically, consider the {\em equivalent classes} resulting from $\preceq$. 
Each equivalent class $E$ for an agent $x$ is a set of agents such that both $y \preceq_x y'$ and $y' \preceq_x y$ for any $y, y' \in E$. 
We only consider maximal equivalent classes, which are not proper subsets of any other equivalent classes, so hereafter by ``equivalent class'' we mean a maximal equivalent class. A feasible way of tie breaking therefore corresponds to specifying a strict order for every equivalent class in $\preceq$. 

Let $\mathcal{E}_x$ be the collection of all equivalent classes in $\preceq$ according to agent $x$'s preference. 
Since $\preceq$ is induced by $p$, for each $E \in \mathcal{E}_x$, we have $v_x(y \given p) = v_x(y' \given p)$ for every $y, y \in E$. 
Equivalently, these relations are defined by a system of $|E| -1 $ equations:
\[
v_x(y_i \given p) - v_x(y_{i+1} \given p) = 0
\]
for $i = 1,\dots, |E| - 1$, where we assume $E = \{y_1, \dots, y_{|E|}\}$ and $y_1 \prec_x \dots \prec_x y_{|E|}$.
More concisely, we write
\[
V_E \cdot p = \mathbf{0}, 
\]
where $V_E$ is a $(|E|-1)$-by-$|E|$ matrix whose $i$-th row is the vector $(v_x(y_i \given \omega) - v_x(y_{i+1} \given \omega))_{\omega \in \Omega}$.

Consider now the matrix $V$ that is a vertical concatenation of all such $V_E$, for all $E \in \mathcal{E}_x$ and all agents $x$.
We have $V \cdot p = \mathbf{0}$ and according to the non-degeneracy assumption, $V$ has at most $|\Omega| - 1$ rows (otherwise, $p = \mathbf{0} \notin \Delta(\Omega)$ would be the only solution to $V \cdot p = \mathbf{0}$), and they are linearly independent.
The following claim then follows.

\paragraph{Claim (2).} {\em 
There exists $p' \in \mathbb{R}^{|\Omega|}$ such that $V \cdot p' < \mathbf{0}$.
}

\smallskip


Indeed, since the row vectors $\mathbf{v}_1,\dots \mathbf{v}_d$ of $V$ are linearly independent, via basic linear algebra it can be proven that for each $i = 1,\dots, h$ there exists a vector $\alpha_i \in \mathbb{R}^{|\Omega|}$ such that $\alpha_i \cdot \mathbf{v}_i = -1$ and $\alpha_i \cdot \mathbf{v}_j = 0$ for all $j \neq i$.
Hence, 
\[
p' := p + \epsilon(\alpha_1 + \dots + \alpha_d)
\]
satisfy the claimed condition for any $\epsilon > 0$ (given that $V \cdot p = \mathbf{0}$).

This means that $v_x(y \given p') < v_x(y' \given p')$ for all $y,y'$ in each of agent $x$'s equivalent classes in $\preceq$ such that $y \prec_x y'$.
By continuity, when $\epsilon$ is sufficiently close to $0$, $p'$ also preserves all the strict orders induced by $p$ (i.e., relations $y \preceq_x y'$ in $\preceq$ such that $y' \not\preceq_x y$).
Hence, $p'$ induces exactly $\prec$. Namely, we have $p' \in C_\prec$.
Now that $p'$ approaches $p$ when $\epsilon$ approaches $0$, it follows immediately that $p \in C_\prec$, which completes the proof.
\end{proof}

\subsection{Bounding the Signal Space by \texorpdfstring{$|\Omega|$}{|Omega|}}
\label{sc:bound-Omega}

Now we have reduced necessary preference profiles.
In our setting, however, this is insufficient because there are still exponentially many possibilities for $M$ in every meta-signal $(\preceq,M) \in \mathcal{L}^\proper \times \mathcal{M}$. 
We need to additional steps to deal with the large space $\mathcal{M}$. 
(On the other hand, it is not feasible to define cells for meta-signals because such cells would intersect at a non-zero measure.)
The following theorem is thus useful for this purpose. 

\begin{restatable}{lemma}{lmmUpperBoundOmega}
\label{lmm:upper-bound-Omega}
There exists an optimal and indicative policy ${\sigma}: \Omega \to \Delta(S)$ such that $S \subseteq \mathcal{L} \times \mathcal{M}$ and $|S| \le |\Omega|$.
\end{restatable}

The above bound has been proven independently in previous work (e.g., \citep[Lemma~4]{aybas2019persuasion}). 
For completeness, we provide a proof in the appendix.

\subsection{Efficient Algorithm for Small \texorpdfstring{$\Omega$}{Omega}}

With the results presented in \Cref{sc:proper-cells,sc:bound-Omega}, we now describe our efficient algorithm for a small $\Omega$.
The main idea of the algorithm is to guess the meta-signal space $S \subseteq \mathcal{L}^\proper \times \mathcal{M}$ of some optimal policy.
Indeed, if $S$ is given, the optimal policy can be computed via the same formulation as LP~\eqref{eq:lp-opt}, where we restrict the meta-signal space to $S$. That is,
\begin{align}
\max\ 
\sum_{\omega \in \Omega} \mu(\omega) \sum_{(\prec, M) \in S} \sigma(\prec, M \given \omega) \cdot u(M \given \omega),
\label{eq:LP-S}
\end{align}
subject to the same constraints as \Cref{eq:lp-opt-1}.



In particular, since by \Cref{lmm:upper-bound-Omega} there are optimal policies supported on a small meta-signal set $S$, the corresponding LP formulation is also small-sized and can be solved efficiently.

Since $\mathcal{M}$ is exponentially large, guessing $S$ directly would be very costly even though we know that $|S| \le |\Omega|$ via \Cref{lmm:upper-bound-Omega}. So, our actual approach is to guess only the preference profiles in $S$, which is tractable because, according to \Cref{lmm:Cpp}, we can restrict our consideration to the space $\mathcal{L}^\proper$.
Formally, we want to guess the multiset $\widehat{\mathcal{L}}$ that contains, {\em with repetition},
a preference profile $\prec$ for every $(\prec, M) \in S$. 
The following observation is key to this approach: 

\begin{observation}
The feasible region of LP~\eqref{eq:LP-S} is independent of the matchings in $S$.
\end{observation}

Given this, our algorithm, presented in \Cref{alg:small-Omega}, proceeds by taking a holistic view over all the vertices of the feasible region, which we know must contain at least one optimal solution irrespective of the objective function.
Thanks to the small size of $S$, the number of vertices grows only polynomially for small $|\Omega|$. 
Each vertex is a partially defined policy, where the probability of each meta-signal and the cell used by this meta-signal are given, but the matching is unspecified. 
The problem boils down to optimizing the following matching problem to pin down the unspecified matching for each vertex.

\begin{definition}[Best matching problem (BMP)]
Given $\prec^1,\dots, \prec^{|\Omega|} \in \mathcal{L}^\proper$ and a number $\gamma(\ell,\omega)$ for each $\ell = 1,\dots,|\Omega|$ such that $\sum_{\ell} \gamma(\ell, \omega) = 1$,
compute matchings $M_1,\dots, M_\ell$ so that the utility yielded by the policy $\sigma$ is maximized while $\sigma$ is stable and indicative, 
where $\sigma(\prec^\ell, M_\ell \given \omega) = \gamma(\ell,\omega)$ for all $\ell$ and $\omega$.
\end{definition}

Note that the choice of each matching in the BMP is independent of each other. Hence, it can be shown that an optimal solution is obtained by picking, for each $\ell$, an arbitrary  
\begin{equation*}
M_\ell \in \argmax_{M \text{ is stable w.r.t. }\prec^\ell} \sum_{\omega} \mu(\omega)\,  \gamma(\ell, \omega)\, 
 u(M \given \omega).\hspace{-1mm}
\end{equation*}
Recall that $u(M \given \omega) = \sum_{(a,b) \in M} u(a,b \given \omega)$, so we can rewrite the objective function as 
$\sum_{(a,b) \in M} \bar{u}(a,b)$, where
$\bar{u}(a,b) := \sum_{\omega} \mu(\omega) \cdot \gamma(\ell, \omega) 
\cdot u(a,b \given \omega)$.
Hence, the optimization problem is essentially the WSM problem with weights given by $\bar{u}$.
When the agents' preferences do not involve ties, as it is the case with $\prec^1, \dots, \prec^{|\Omega|}$, the WSM problem admits a polynomial-time algorithm.\footnote{It is intractable in general when there might be ties. Recall \Cref{thm:nph}.}
In turn, the BMP can be solved efficiently.


\begin{restatable}{lemma}{lmmBMPPolyTime}
\label{lmm:BMP-poly-time}
The BMP can be solved in time $\poly(|\Omega|,n)$.
\end{restatable}

\begin{proof}
Note that the choice of each matching is independent of each other. Hence, an optimal solution to the BMP is obtained by picking, for each $\ell$, an arbitrary  
\begin{equation}
\label{eq:BMP-solution}
\hspace{-2mm}
M_\ell \in \argmax_{M \text{stable w.r.t. }\prec^\ell} \sum_{\omega} \mu(\omega)\,  \gamma(\ell, \omega)\, 
 u(M \given \omega).\hspace{-1mm}
\end{equation}
Indeed, the stability of $\sigma$ requires $M_\ell$ being stable with respect to $\Pro(\cdot \given \prec^\ell, M_\ell)$, while the indicativeness of $\sigma$ requires $\Pro(\cdot \given \prec^\ell, M_\ell) \in C_{\prec^\ell}$. Hence, the set of matchings that are stable with respect to $\prec^\ell$ contains exactly the feasible choices of $M_\ell$.

Note that $u(M \given \omega) = \sum_{(a,b) \in M} u(a,b \given \omega)$, so we can rewrite the objective function in \Cref{eq:BMP-solution} as 
$\sum_{(a,b) \in M} \bar{u}(a,b)$, where
\[
\bar{u}(a,b) := \sum_{\omega} \mu(\omega) \cdot \gamma(\ell, \omega) 
\cdot u(a,b \given \omega).
\]
Therefore, the optimization problem in \Cref{eq:BMP-solution} is essentially the WSM problem,
where we are asked to find a stable matching with the maximum weight with respect to the weights given by $\bar{u}$. 
When the agents' preferences do not involve ties (as it is the case with $\prec^\ell$), the problem is known to be tractable via an LP formulation \cite{vate1989linear}. The time complexity then follows.
\end{proof}

\begin{algorithm}[t]
\begin{itemize}[leftmargin=0mm]
\item 
For each multiset $\widehat{\mathcal{L}} = \{\prec^1, \dots, \prec^{|\Omega|} \} \subseteq \mathcal{L}^\proper$ do:

\begin{enumerate}[leftmargin=7mm]

\item 
Compute the set $\Gamma$ of vertices of the feasible region of LP~\eqref{eq:LP-S}, where we let 
$S = \{(\prec^\ell, M_\ell)\}_{\ell=1,\dots,|\Omega|}$ 
for arbitrarily chosen $M_1, \dots, M_{|\Omega|}$;

\item 
For each $\gamma \in \Gamma$, solve the BMP for a set of new matchings $M_1,\dots, M_{|\Omega|}$.
\end{enumerate}

\item 
Pick the $\widehat{\mathcal{L}}$ and $\gamma$ that lead to the maximum utility in Step 2. 

\item 
Output $\sigma$ such that $\sigma(\prec^\ell, M_\ell \given \omega) = \gamma(\ell, \omega)$ for every $\ell,\omega$.
\end{itemize}

\caption{Compute an optimal policy (when $\Omega$ is small)\label{alg:small-Omega}}
\end{algorithm}
Summarizing the above results, \Cref{alg:small-Omega} computes an optimal policy. The following time complexity of the algorithm can be established readily.

\begin{restatable}{theorem}{thmPublicSmallWorldsPoly}
\label{thm:public-small-worlds-poly}
An optimal policy can be computed in time $\poly(n)$ for non-degenerate instances when $|\Omega|$ is small.
\end{restatable}

\begin{proof}
The result follows by analyzing the time complexity of \Cref{alg:small-Omega}.
Specifically, there are $|\mathcal{L}^+|^{|\Omega|} = O(n^{3|\Omega|^2})$ multisets to enumerate, 
where $|\mathcal{L}^\proper|$ is bounded by \Cref{lmm:num-proper-cells}.
The number of vertices each $\Gamma$ has is at most ${2 n^3 |\Omega| \choose |\Omega|^2} = O((2n^3|\Omega|)^{|\Omega|^2})$, where $2 n^3 |\Omega|$ is an upper bound on the number of constraints of LP~\eqref{eq:LP-S} and $|\Omega|^2$ is the number of variables.
Summing up the numbers as well as the time complexity for solving \Cref{lmm:BMP-poly-time} gives the stated complexity. 
\end{proof}

As a final remark, one may wonder if the non-degeneracy assumption alone is already sufficient for the tractability of the problem. 
We prove that this is not the case via a reduction from the standard public persuasion problem, which is known to be intractable.

\begin{restatable}{theorem}{thmArbitraryNonDegenerateHardness}
\label{thm:arbitrary-non-degenerate-hardness}
It is NP-hard to compute an optimal policy even for non-degenerate instances (when $|\Omega|$ is not fixed).
\end{restatable}

\begin{proof}
We reduce the problem of computing an optimal public policy to the following {\em public persuasion} problem. In the public persuasion problem, every agent $i \in I$ is free to choose an action $j \in J$ to perform. Each action $j$ generates a payoff $v_i(j, \omega)$ for the agent themselves as well as a payoff $u(j, \omega)$ for the principal, where $\omega \sim \mu$ is the world, whose realization is known only to the principal. 
Same as our model, the principal sends a public signal to inform the agents about the observed world, and then every agent picks an action that is optimal with respect to their posterior belief induced by the public signal.
The goal is to design a policy that maximizes the principal's expected payoff. 
This problem is known to be NP-Hard [{\citealp[Theorem 6.2]{dughmi2017algorithmic}}] and their reduction still applies when small random noises are added to the agents' values in the reduced instance, which amounts to the non-degenerate case.

The reduction is as follows. Let $\langle I, J, \Omega, \mu, v \rangle$  be an instance of the above public persuasion problem, where $J$ contains only two actions.
We construct an instance of our problem, where: 
$A = I \cup A_{\mathrm{dummy}}$ contains all agents in $I$ as well as a set $A_{\mathrm{dummy}}$ of dummy agents; and $B$ contains $|I|$ copies of the set $J$, each corresponding to an agent in $I$.
Hence, the size of $B$ is $2|I|$. We let the size of the dummy set $A_{\mathrm{dummy}}$ be $|I|$ so that $A$ and $B$ have the same size.
Intuitively, matching an agent in $A$ with an agent in $B$ corresponds to matching an agent in $I$ to an action in $J$ in the original problem we reduce from.
We then set the payoff functions of agents in $I$ to be the same as those in the original instance, and set payoff functions of all the other agents (including the dummies in $A_{\mathrm{dummy}}$ and all agents in $B$).

It is straightforward that a matching between $A$ and $B$ is stable in the above instance if and only if the corresponding matching between $I$ and $J$ is a matching between the agents with their optimal actions. Since the preferences of the dummy agents are strict and do not depend on the world, we can add small noise to their preferences, without actually changing their preferences under any posterior distribution, and so the resulting instance is non-degenerate.
\end{proof}

\section{Using Private Signals}
\label{sc:private}

We now turn to private persuasion.
As a generalization of public persuasion, private persuasion inherits the general hardness, and may lead to strict utility improvement. We then consider the special cases as in the previous sections.

\paragraph{Private Persuasion with Small \texorpdfstring{$T$}{T}}

As a main challenge in the private case, agents may receive different signals and derive different posteriors about the world. In the type-based model, one may conjecture that there is no need to differentiate agents of the same type, and it is without loss of optimality to send the same signal to the same type. 
It turns out that this is not true, as is proven by the next example.

\begin{example}
\label{exp:cannot-signal-by-type}
Let the types be $\mathcal{A} = \{A_1, A_2\}$ and $\mathcal{B} = \{B_1, B_2, B_3, B_4\}$. There are four agents in $A_1$, two in each of $A_2, B_1$, and $B_2$, and one in each of $B_3$ and $B_4$.
The possible worlds are $\Omega = \{\omega_1, \omega_2\}$, with prior $\mu(\omega_1) = \mu(\omega_2) = 0.5$.
For all $\omega \in \Omega$,
The principal's utility is: $u(A_i, B_j \given \omega) = 1$ only if $i = 1$ and $j \in \{1, 2\}$; otherwise, $u(A_i, B_j \given \omega) = 0$.
The agents' values are given in \Cref{tab:cannot-signal-by-type}.
\end{example}

\begin{table}[h]
\centering
\small
\begin{tabular}{@{\extracolsep{4pt}}rrrrrrrrr}
\toprule
      & \multicolumn{2}{c}{$B_1$} & \multicolumn{2}{c}{$B_2$} & \multicolumn{2}{c}{$B_3$} & \multicolumn{2}{c}{$B_4$} \\ \cline{2-3}\cline{4-5}\cline{6-7}\cline{8-9}
      & $\omega_1$  & $\omega_2$  & $\omega_1$  & $\omega_2$  & $\omega_1$  & $\omega_2$ & $\omega_1$  & $\omega_2$ \\ \midrule
$A_1$ & $-1$ & $3$ & $0$ & $0$ & $2$ & $-1$ & $0$ & $0$ \\
$A_2$ & $0$ & $0$ & $0$ & $0$ & $0$ & $0$ & $0$ & $0$
    \\     
\bottomrule
\end{tabular}
\qquad\quad
\begin{tabular}{@{\extracolsep{4pt}}rrrrrrrrr}
\toprule
      & \multicolumn{2}{c}{$A_1$} & \multicolumn{2}{c}{$A_2$} \\ \cline{2-3}\cline{4-5}
      & $\omega_1$  & $\omega_2$  & $\omega_1$  & $\omega_2$  \\ \midrule
$B_1$ & $1$ & $1$ & $0$ & $0$ \\
$B_2$ & $1$ & $1$ & $0$ & $0$ \\
$B_3$ & $-1$ & $2$ & $0$ & $0$ \\
$B_4$ & $1$ & $1$ & $0$ & $0$
    \\     
\bottomrule
\end{tabular}
\vspace{5mm}
\caption{Agents' values for \Cref{exp:cannot-signal-by-type}. \label{tab:cannot-signal-by-type}}
\end{table}

In this example, the matching with the highest value for the principal, irrespective of the world, is $M$ with 
$M(A_1, B_1) = M(A_1, B_2) = 2$ and $M(A_2, B_3) = M(A_2, B_4) = 1$, which yields a utility of $4$. No other matchings can achieve this utility.
To make this matching stable, the principal must always reveal the world to agents in $B_3$, as well as agents in $A_1$ who are matched to $B_2$. 
However, the agents from $A_1$ that are matched to $B_1$ must not know the world, otherwise the matching will not be stable under $\omega_1$.

The example also implies a strict utility improvement by using private signals compared to the public case, since public policies do not differentiate among any agents at all.

Nevertheless, we prove that it suffices to differentiate the agents at a {\em subtype} level. 

A subtype (\Cref{def:subtype}) includes both the agent's own type and the type of their partner in a given matching.
This notion is essential because otherwise it would be unclear how a policy could even be represented efficiently if agents need to be differentiated at the individual level, given that the number of agents may grow exponentially in the type-based model.
In what follows, when a private policy sends meta-signal $(\preceq, M)$, each agent $x$ receives $(\preceq_x, M)$. 

\begin{definition}[Subtype]
\label{def:subtype}
Given a matching $M$, the {\em subtype} of an agent $x$ is the pair $(s, t)$ such that $s,t \in \mathcal{A} \cup \mathcal{B}$ and $x \in s, M(x) \in t$.
\end{definition}

\begin{restatable}{theorem}{thmPrivateOptimalIndicativeUsesV}
\label{thm:private-optimal-indicative-uses-V}
There exists an optimal private policy $\sigma: \Omega \to \Delta(S)$, $S \subseteq \mathcal{L} \times \mathcal{V}^*$, that is also indicative (recall that $\mathcal{V}^* = \bigcup_{P \in \mathcal{A}\times\mathcal{B}} \mathcal{V}_P$ is defined in \Cref{sc:small-types}). 
Moreover, it holds for every $(\preceq, M) \in S$ that:
$\preceq_x = \preceq_{x'}$ for every pair of agents $x$ and $x'$ of the same subtype in $M$.
\end{restatable}

\begin{proof}
The fact that it suffices to use signals in $\mathcal{L}$ follows the same argument as \Cref{thm:rp-C}: we can always merge signals inducing the same preference without changing the stability of the policy as well as the utility it yields. Hence, we can assume that there exists an optimal policy $\sigma: \Omega \to \Delta(\mathcal{L} \times \mathcal{M})$ that is indicative.

Next, we show that, similarly to the public signal case, we have the flexibility to change each matching in the policy to any other matching that has the same prototype. 
For better intuition, it would be convenient to view the signaling process under a policy $\sigma$ as one where: 
a matching is selected first, with probability $\sigma(M \given \omega) = \sum_{\preceq \in \mathcal{L}} \sigma(\preceq, M \given \omega)$ for each $M$; and then a signal $\preceq$ is selected, with probability $\sigma(\preceq \given M, \omega)$.  

We first modify the second phase of the process and change the conditional probability $\sigma(\preceq \given M, \omega)$ in a way such that agents of the same subtype always receive the same signal.
Specifically, for each subtype $\tau$ in $M$, we pick an arbitrary agent of subtype $\tau$ as the representative of this subtype.
The choices define a map $f: \mathcal{L} \to \mathcal{L}$, such that whenever some $\preceq$ is selected by $\sigma$, $f$ maps it to $\preceq'$ such that 
\[
\preceq'_x = \preceq_\tau
\]
for every agent $x$ of subtype $\tau$,
where $\preceq_\tau$ is the preference of the representative agent of subtype $\tau$.
Hence, if the policy always sends $\preceq'$ when originally it was to send $\preceq$, all agents of subtype $\tau$ will acquire the same belief $\mathbb{P}(\omega|\preceq_\tau, M) = \frac{\mu(\omega) \cdot \sigma(\preceq_\tau, M \given \omega)} {\sum_{\omega'} \mu(\omega') \sigma(\preceq_\tau, M \given \omega')}$ as the representative agent.
As a result, the policy remains indicative and $M$ remains stable with respect to the posterior beliefs induced by every meta-signal containing $M$, while now agents of the same subtype always receive the same meta-signal.
Note that the modification does not change the marginal probability $\sigma(M \given \omega)$, so the utility the policy yields for the principal remains the same. 

Now that we can change the agents' posterior beliefs without breaking stability and indicativeness of the policy, the same approach also allows us to modify every matching to any other matching with the same prototype. 
This will cause changes to the subtypes of some agents, but it suffices to choose $\preceq'_x = \preceq_{\tau'}$ for the subtype $\tau'$ of $x$ in $M'$, instead of $M$, when defining $f$.
Note that it may happen that multiple meta-signals will be modified to the same $(\preceq', M')$, which will cause, for each agent, a merge of the posterior beliefs that were to be induced if we also add a tag in $(\preceq', M')$ to help the agent to identify the original meta-signal $(\preceq', M')$ is modified from. However, since $(\preceq', M')$ still induces $\preceq'$, this will not change the stability of $M'$.

Therefore, same as the public case, we can further change $M$ to an arbitrary optimal solution $M'$ to the following problem without compromising the utility:
\begin{equation*}
\max_{M' \in \mathcal{M}_{P'}, P' \subseteq P}\quad
\sum_{\omega \in \Omega} \mu(\omega) \cdot \sigma(M \given \omega) \cdot u(M' \given \omega).
\end{equation*}
The optimization takes the same form as \Cref{eq:opt-M-prototype} but with the marginal probabilities $\sigma(M \given \omega)$ in the objective since we now modify altogether all the meta-signals that contain $M$.
In the LP form, the feasible region of the problem is the same as the one defined by \Crefrange{eq:lp-prototype-1}{eq:lp-prototype-4}. Hence, for the same reason as in the public case, it suffices to consider matchings in $\mathcal{V}^*$, which are defined by the vertices of the feasible regions.
The stated result then follows.
\end{proof}

Based on \Cref{thm:private-optimal-indicative-uses-V}, \Cref{alg:small-T} then applies to the private case, too, a slight difference being that the meta-signal $(\preceq, M)$ in \Cref{eq:lp-opt-1} should be modified to the one $(\preceq_x, M)$ which agent $x$ sees.
The same analysis as \Cref{thm:type-model-time-complexity} then gives the same time complexity result for the private case. We state the result below and omit the proof.

\begin{theorem}
\label{thm:P-small-types-private}
An optimal private policy can be computed in $\poly(|\Omega|, \log n)$ time when $T$ is a constant.
\end{theorem}

\paragraph{Private Persuasion with Small \texorpdfstring{$\Omega$}{Omega}}

Unfortunately, the case of small $|\Omega|$ does not admit any efficient algorithm under private signals, even when there are only two possible worlds in $\Omega$. We present the hardness result next. The hard instances in our reduction have a positive measure, so excluding degenerate instances in a zero-measure set will not reduce the complexity. Our reduction is based on the intractability of the maximum size stable matching problem with ties and incomplete lists, in a special case where: there is at most one tie in the preference order of every agent in $A$ and no ties in the orders of agents in $B$, every tie involves two agents, and every agent appears in at most one tie. 
The hardness of this special case, which we prove in Appendix, may also be of independent interest.

\begin{restatable}{lemma}{lmmNphSMTISpecialCase}
\label{lmm:nph-SMTI-special-case}
The problem of maximizing the size of stable matching with ties and incomplete lists (MAX-SMTI) is NP-hard, even when each tie occurs at the end of some agent’s preference list, ties occur on one side only, each tie is of length two, and every agent appears in one tie or less.
\end{restatable}

\begin{restatable}{theorem}{thmNphSmallWorldsPrivate}
\label{thm:nph-small-worlds-private}
It is NP-hard to compute an optimal private policy even when $|\Omega|=2$, and the hard instances have a nonzero measure.
\end{restatable}

\section{Conclusion}
\label{sc:conclusion}

We have studied the problem of persuading stable matching, which combines persuasion and stable matching formation.
While the problem is generally hard, we demonstrated that some reasonable relaxations exist for which the problem is tractable. We believe that these relaxations and methods could be used for more general cases than matching persuasion, and plan to investigate further in this direction in future work.
We also presented general results on the hardness of stable matching, pushing further some currently known hard sub-cases.

One can also consider other variants of private persuasion in our model. In some scenarios, a pair of agents matched together may communicate signals they received, which would further change their beliefs and influence the stability of the matching. In some other scenarios, the principal may choose to hide her selection of $M$ from the agents, perhaps only informing each agent privately who they are to be matched with. In this case, each agent would establish a probabilistic belief about the overall matching. The stability notion would need to be redefined to adapt to such probabilistic beliefs. 

\section{Acknowledgments}

This research has been partially supported by the Israel Science Foundation under grant 2544/24.


\bibliographystyle{cas-model2-names}

\bibliography{cas-refs}

\clearpage

\appendix

\section{Omitted Proofs}

\lmmUpperBoundOmega*

\begin{proof}
By \Cref{lmm:upper-bound-Omega}, we know that an optimal indicative policy ${\sigma}: \Omega \to \Delta(S)$ exists, such that $S \subseteq \mathcal{L} \times \mathcal{M}$ and $\Pro(\cdot \given \preceq, M) \in C_\preceq$ for every $(\preceq, M) \in S$

Given $\sigma$, the prior distribution $\mu$ can be written as a convex combination of posterior distributions induced by the meta-signals:
\begin{align*}
\mu(\cdot) = \sum_{s \in S} \Pro(s) \cdot \Pro(\cdot \given s),
\end{align*}
where $\Pro$ is the probability measure induced by $\sigma$, and $\Pro(s) = \sum_{\omega' \in \Omega} \mu(\omega') \cdot \sigma(s \given \omega')$ is the marginal probability of $s$ being sent.
We can also write the principal's expected utility as
\begin{align}
\label{eq:utility-decompose}
U(\sigma) 
&= \sum_{s \in S} \Pro(s) \cdot \underbrace{ \sum_{\omega \in \Omega} \Pro(\omega \given s) \cdot u(s \given \omega)}_{:= V(s)},
\end{align}
where for any $s = (\preceq, M) \in S$, we define $u(s \given \omega) = u(M, \omega)$. 
For ease of description, we denote the result of the inner summation as $V(s)$.

Now consider the following linear program (LP).
\begin{align*}
\max_{\varphi} 
\quad & \sum_{s \in S} \varphi(s) \cdot V(s) \\
\text{subject to}  \quad
&\mu(\omega) = \sum_{s \in S} \varphi(s) \cdot \Pro(\omega \given s) & \text{for } \omega \in \Omega \\
&\varphi(s) \ge 0 & \text{for } s \in S \\
&\sum_{s \in S} \varphi(s) = 1
\end{align*}
Clearly, $\varphi(s) = \Pro(s)$ for all $s \in S$ is a feasible solution to this LP.
In what follows, we will argue that the following claims holds:
\begin{itemize}
\item[(1)] There exists an optimal solution $\varphi^*$ to the above LP whose support size is at most $|\Omega|$.
\item[(2)] $\varphi^*$ corresponds to a stable and indicative policy that yields at least as much utility for the principal as $\sigma$ does.
\end{itemize}
The latter gives the desired result.

\paragraph{Claim (1).}
Suppose that $\varphi'$ is an optimal solution supported on $S' \subseteq S$. 
It suffices to show that if $|S'| > |\Omega|$ then there exists another optimal solution supported on $S''$ and $|S''| \le |S'| - 1$.

Since $\varphi'$ is optimal, the KKT conditions hold.
Let $\alpha(\omega)$, $\beta(s)$, and $\gamma$ be the dual variables corresponding to three sets of constraints in the LP.
Since $\varphi'(s) > 0$ for all $s$ in the support of $\varphi'$, by complementary slackness in the KKT conditions, we have $\beta(s) = 0$ for all $s \in S'$.
Moreover, by stationarity, the following equation holds for all $s \in S$:
\[
\sum_{\omega \in \Omega} \Pro(\omega \given s) \cdot \alpha(\omega) - \sum_{s \in S} \beta(s) + \gamma = V(s).
\]
In particular, for those $s \in S'$, since $\beta(s) = 0$, we have
\[
\sum_{\omega \in \Omega} \Pro(\omega \given s) \cdot \alpha(\omega) + \gamma = V(s).
\]
Viewing $\alpha$ and $\gamma$ as variables, this gives a system of $|S'|$ linear equations, with $|\Omega| + 1$ variables.
In the matrix form, this can be written as
$
\begin{pmatrix}
\mu & \mathbf{1}
\end{pmatrix}
\begin{pmatrix}
\alpha \\ \gamma 
\end{pmatrix} = V
$
where $\mu$ denotes a $|S'|$-by-$|\Omega|$ matrix, $\mathbf{1} = (1,\dots,1)^\mathsf{T}$, $\alpha$ is a size-$|\Omega|$ column vector, and $V$ is a size-$|S'|$ column vector.
Note that $\begin{pmatrix}
\mu & \mathbf{1}
\end{pmatrix}$ has rank at most $|\Omega|$ as $\mathbf{1}$ can be expressed as the sum of the columns of $\mu$.
Since the system has at least one solution, the rank of $\begin{pmatrix}
\mu & \mathbf{1} \ |\ V
\end{pmatrix}$
must be at most $|\Omega|$, too.
Now that there are $|S'|$ rows in $\begin{pmatrix}
\mu & \mathbf{1} \ |\ V
\end{pmatrix}$ and by assumption $|S'| > |\Omega|$, 
it must be that we can write one row as the linear combination of all the other rows.
Namely, for some $t \in S'$, there exists a number $c_s$ for each $s \in S' \setminus \{t\}$ which makes the following equations hold:
\begin{subequations}
\begin{align}
& \sum_{s \in S' \setminus \{t\}} \Pro(\omega \given s) \cdot c_s = \Pro(\omega \given t) \quad \text{ for all } \omega \in \Omega \label{eq:linear-combin-a}\\
& \sum_{s \in S' \setminus \{t\}} c_s = 1 \label{eq:linear-combin-b}\\
& \sum_{s \in S' \setminus \{t\}} V( s) \cdot c_s = V( t) \quad \text{ for all } \omega \in \Omega \label{eq:linear-combin-c}
\end{align}
\end{subequations}

Construct the following solution $\varphi''$ to the LP.
Let
\begin{align*}
&\varphi''(s) = 0 \quad \text{for all } s \notin S', \\
&\varphi''(t) = (1 - \lambda) \cdot \varphi'(t), \\
\text{ and } \quad
&\varphi''(s) = \varphi'(s) + \lambda \cdot c_s \cdot \varphi'(t)
\quad \text{for all } s \in S' \setminus \{t\},
\end{align*}
where $\lambda \in [0,1]$ is a parameter to be determined shortly.
Given \Cref{eq:linear-combin-c}, it can be verified that $\varphi''$ yields the same objective value as $\varphi'$ does, irrespective of the choice of $\lambda$:
\begin{align*}
\sum_{s \in S} \varphi''(s) \cdot V(s) 
&= \sum_{s \in S'} \varphi''(s) \cdot V(s) \\
&= (1 - \lambda) \cdot \varphi'(t) \cdot V(t) +
\sum_{s \in S' \setminus \{t\}} \left(\varphi'(s) + \lambda \cdot c_s \cdot \varphi'(t) \right) \cdot V(s) \\
&= \sum_{s \in S'} \varphi'(s) \cdot V(s) \\ 
&= \sum_{s \in S} \varphi'(s) \cdot V(s).
\end{align*}
Similarly, using \Cref{eq:linear-combin-a,eq:linear-combin-b}, we can verify that $\varphi''$ also satisfies the first and the third constraints of the LP.
The only thing missing is the second constraint: $\varphi(s) \ge 0$ for all $s\in S$.
Indeed, this can be ensured by choosing a sufficiently small $\lambda$, so we obtain $\varphi''$ as another optimal solution to the LP.

Besides that, if we start by letting $\lambda = 0$ and gradually increase it, we can expect that at some point either $\varphi''(s)$ becomes $0$ for some $s \in S' \setminus \{t\}$ (note that $c_s$ might be negative), or $\varphi''(t)$ becomes $0$ (i.e., when $\lambda$ reaches $1$). In either case, at least one more component of $\varphi''$ will become zero, so $\varphi''$ is supported on a strictly smaller set.

\paragraph{Claim (2).}

To convert $\varphi^*$ into a stable policy $\tilde{\sigma}$, we set 
\[
\tilde{\sigma} ( s \given \omega) = \frac{\varphi^*(s) \cdot \Pro( \omega \given s)} {\mu(\omega)}
\]
for every $s \in \widetilde{S}$ and $\omega \in \Omega$.
Note that $\Pro$ still denotes the probability measure induced by $\sigma$.

We can first verify that $\tilde{\sigma} ( \cdot \given \omega) \in \Delta(\widetilde{S})$ is a valid distribution. 
Indeed, we have $\tilde{\sigma} ( s \given \omega) \ge 0$ and 
\begin{align*}
\sum_{s \in \widetilde{S}} \tilde{\sigma} ( s \given \omega) 
= \frac{ \sum_{s \in \widetilde{S}} \varphi^*(s) \cdot \Pro( \omega \given s)} {\mu(\omega)} 
= 1,
\end{align*}
where the last equality follows by the first constraint in the LP.

Moreover, the posterior distribution $\widetilde{\Pro}( \cdot \given s)$ induced by every signal $s \in \widetilde{S}$ under $\tilde{\sigma}$ is the same as that under $\sigma$ since
\begin{align*}
\widetilde{\Pro}( \omega \given s) 
&= \frac{\mu(\omega) \cdot \tilde{\sigma}(s \given \omega)} {\sum_{\omega' \in \Omega} \mu(\omega') \cdot \tilde{\sigma}(s \given \omega')} \\
&= \frac{\varphi^*(s) \cdot \Pro( \omega \given s)} {\sum_{\omega' \in \Omega} \varphi^*(s) \cdot \Pro( \omega' \given s)}
= \Pro( \omega \given s).
\end{align*}
Hence, $\tilde{\sigma}$ is stable and indicative. In addition, same as in \Cref{eq:utility-decompose}, we can now write the principal's expected utility under $\tilde{\sigma}$ as $U(\tilde{\sigma}) = \sum_{s \in S} \varphi^*(s) \sum_{\omega} \Pro(\omega \given s) \cdot u(s \given \omega)$.
This is exactly the objective value of the LP above for the solution $\varphi^*$. Since $\varphi^*$ is the optimal, we then have 
$U(\tilde{\sigma}) \ge \sum_{s \in S} \sum_{\omega} \Pro(\omega \given s) \cdot u(s \given \omega) = U(\sigma)$.
\end{proof}

\lmmNphSMTISpecialCase*


\begin{proof}
Given an instance of the MAX-SMTI problem, $\mathcal{M} = (A, B, p, t)$:

Two sets $A, B$ for the agents on each side.
A (possibly incomplete) preferences list $p_c$ for each agent $c \in A \cup B$ by which he ranks the agents on the other side, along with a possible pair $t_a$ that may contain two or zero agents for each $a \in A$ ($t_a \cap p_a = \emptyset$).

We will construct the following instance, $\mathcal{M}' = (A', B', p', t')$:

\begin{align*}
    & B'' = \{b \in B | \nexists a \in A : b \in t_a \} 
    & \\
    & B''' = \{b^j_i | b_i \in B, a_j \in A, b_i \in t_{a_j} \}
    & \\
    & B'''' = \{\tilde{b}^j_i | b_i \in B, a_j \in A, b_i \in t_{a_j} \}
    & \\
    & B' =  B'' \cup B''' \cup B''''
    & \\
    & A'' = \{a^j_i | b_i \in B, a_j \in A, b_i \in t_{a_j} \} 
    & \\
    & A''' =  \{\tilde{a}^j_i | b_i \in B, 1 \le j \le |\{a \in A | b \in t_a \}| - 1\}
    & \\
    & A' = A \cup A'' \cup A'''
\end{align*}

\smallskip


Basically, each agent in $b \in B$ that appears in $n>1$ ties is duplicated into $n$ different agents in $B''$, for which we define additional $n$ dummy agents in both $A'', B''''$ and another $n-1$ dummy agents in $A'''$. As will be defined shortly, the preferences of the agents are such that in every maximum-size stable policy, $n-1$ of the agents from $B'''$ will be matched to $n-1$ of the corresponding agents from $A''$ defined for $b$. The agent from $B''$ that won't be matched to agents from $A''$ will serve an identical role to that of $b$ in the original instance.

The preferences are defined by:
\begin{align*}
    & p'_{b_i} : b_i \in B'' = p_{b_i} 
    & \\
    & p'_{b^j_i} : b^j_i \in B''' = (a^j_i) \oplus p_{b_i} 
    & \\
    & p'_a : a \in A = \bigoplus_{b_k=p_{a_j}: j=1}^{dim(p_a)} \begin{cases} 
    (b_k) & b_k \in B'' \\
    (b^l_k | b^l_k \in B''')^{\inf}_{l=1} & b_k \not\in B''
    \end{cases}
    & \\
    & p'_{a^j_i} : a^j_i \in A'' = (\tilde{b}^j_i, b^j_i)
    & \\
    & p'_{\tilde{b}^j_i} : \tilde{b}^j_i \in B'''' = (\tilde{a}^l_i | \tilde{a}^l_i \in A''')^{\inf}_{l=1} \oplus (a^j_i)
    & \\
    &  p'_{\tilde{a}^j_i} : \tilde{a}^j_i \in A''' = \emptyset
\end{align*}
And the ties:
\begin{align*}
    & t'_{a_i} : a_i \in A = \{b^i_j | b_j \in t_{a_i}\} \\
    & t'_{a^j_i} : a^j_i \in A'' =  \emptyset \\
    & t'_{\Tilde{a}^j_i} :\Tilde{a}^j_i \in A''' =  \{\tilde{b}^j_i, \tilde{b}^{j + 1}_i \}
\end{align*}

It is easy to see that in the new instance, every agent appears in one tie list or doesn't appear at tie list at all.

Let us assume a matching $m$ is stable under $\mathcal{M}$; We will show that there exist a matching $m'$ that is stable under $\mathcal{M}'$, where $|m'| = |m| + |A''| + |A'''|$.

Correspondingly to the original matching, we define:
\begin{align*}
    & m^1 = \{(a, b) | (a, b) \in m : \nexists a' \in A, b \in t_{a'} \}
    & \\
    & m^2 = \{(a_i, b^i_j) | (a_i, b_j) \in m: b_j \in t_{a_i} \}
    & \\
    & m^3 = \{(a_i, b^k_j) | (a_i, b_j) \in m: \exists a_k \in A, b_j \in t_{a_k} \land b_j \not\in t_{a_i} \}
\end{align*}
And in order to handle the dummy agents and ensure stability and maximum size:
\begin{align*}
&m^4 = \{ (a^j_i, b^j_i) \,|\, b^j_i \in B''' : \not\exists a : (a, b_j^i) \in m^2 \cup m^3 \\
&\qquad\qquad\qquad 
\land (\exists a \in A: (a, b_j) \in m \lor \exists k > j : b^k_i \in B''')\}
\\
& m^5 = \{(a^j_i, \tilde{b}^j_i) \,|\, a^j_i \in A'' : \not\exists b : (a^j_i, b) \in m^4 \}
& \\
& m^6 = \bigcup_i zip \left( \left(\tilde{a}^j_i \in A''' \right)^{\inf}_{j=1}, \left(\tilde{b}^j_i \in B'''' \,|\, \nexists a : (a, \tilde{b}^j_i) \in m^5 \right)^{\inf}_{j=1} \right)
\end{align*}
Moreover, $m' =  \bigcup_{i=1}^6 m^i$.
(A $zip$ of two tuples $(a_1, a_2, ..., a_n)$, $(b_1, b_2, ..., b_n)$ is the set $\{(a_i, b_i)| 1 \le i \le n \}$)

It is easy to see that for every pair in $m$ we have a pair in $m'$, and a closer look reveals also that every agent in $A'', A'''$ is matched under $m'$, and so $|m'| = |m| + |A''| + |A'''|$.

Let us assume that $m'$ is not stable, and so there exist $a \in A', b \in B'$ that prefer each other over their matched partners.
\begin{itemize}
\item
If $a = \Tilde{a}^j_i \in A'''$: Since $a$ have only a tie, without preferences, he doesn't prefer any agent over another. In addition, it can be seen that $\Tilde{a}^j_i$ is matched to either $b^j_i$ or $b^{j+1}_i$, both belong to his tie group.

\item
If $a = a^j_i \in A''$: Since $a$ have only two agents in his preferences list, and all agents in $A''$ are matched, it must be that $b \in B''''$. However, it can't be that such $b$ prefers $a$, since all agents from $B''''$ are matched and there is at most one agent from $A''$ in their preferences lists, which is always the last.

\item
If $a_i \in A, b = b^j_k \in B'''$: If $m'(b^j_k) \in A''$, $b^j_k$ won't prefer $a_i$ over his current partner and so it must be that $m'(b^j_k) = a_j \in A$. But in this case $m(b_k) = a_j$ and it can be seen that $b_k$ must prefer $a_i$ under the original instance too, and so $m$ is not stable.

\item
If $a = a_i \in A, b = b_j \in B''$: It can be seen that $m'(b_j) \in A$. If $m'(a_i) \in B'' \subseteq B$, $m$ is unstable too, since the order of the agents from $A, B''$ are the same under both systems; Otherwise, $m'(a_i) = b^i_k \in B'''$, and then $m(a_i)=b_k$. But it can be seen that since all the $b^i_k$ are in a row and in equivalent relation to the other agents as in $\mathcal{M}$, $m$ is still unstable.
\end{itemize}

Since for every possible combination we got to a contradiction to the stability of $m$, $m'$ must be stable too.

Now we will see the other direction. Given a maximum-size stable matching $m'$ under $\mathcal{M}'$, we will construct a stable matching $m$ under $\mathcal{M}$ such that $|m| = |m'| - |A''| - |A'''|$.


To begin with, we will note that all agents in $A'''$ are matched to agents in $B''''$ under $m'$: otherwise the matching can be extended by matching more agents of these types together, in contradiction to the optimality assumption, or that these agents from $B''''$ are matched to other agents which contradicts the stability of $m'$. Then, there is exactly one agent in $B''''$ for each tie that isn't matched to agents in $A'''$. By a similar argument we can see that this agent is matched to some agent in $A''$, and that the rest of the agents in $A''$ are all matched to agents in $B''$, such that again for each tie one agent isn't matched to agents in $A''$. By this we can see that all agents in $A'', A'''$ are matched under $m'$.

We define $m$ as follows:

\begin{align*}
    & m'' = \{(a, b) \in m' | a \in A, b \in B''\}
    & \\
    & m''' = \{(a, b_i) \in m' | a \in A, b^j_i \in B'''\}
    & \\
    & m = m'' \cup m'''
\end{align*}

Since every agent from $A''$ and $ A'''$ is matched under $m'$, and every agent in $A$ that is matched under $m'$ is also matched under $m$, we can see that $|m| + |A''| + |A'''| = |m'|$.

Let us assume that $m$ is not stable under $\mathcal{M}$, and so there are agents $a_i \in A, b_j \in B$ that prefer each other over their matched partners. 
\begin{itemize}
\item 
If $b_j \in B''$ then $p'_{b_j} = p_{b_j}$ and $m(b_j) = m'(b_j)$ and so $a_i, b_j$ prefer each other under $\mathcal{M}'$ too, in contradiction to the stability assumption.

\item 
If $b_j \not\in B''$, then there exists $b^k_j \in B'''$ such that $m'(b^k_j)=a_k$ or, not exists at all if $m(b_j)$ also not exists. Then, we can see that $a_i, b^k_j$ will prefer each other under $\mathcal{M}'$ too, in contradiction to the assumption.
\end{itemize}
Hence, the stability of $m$ under $\mathcal{M}$ follows from the stability of $m'$ under $\mathcal{M}'$.

We have shown that for every stable matching $m$ under $\mathcal{M}$ there exists a stable matching $m'$ under $\mathcal{M}'$ such that $|m| + |A''| + |A'''| = |m'|$. In addition, for every maximum-size stable matching $m'$ under $\mathcal{M}'$ there exists a stable matching $m$ under $\mathcal{M}$ such that $|m| + |A''| + |A'''| = |m'|$; Combining these two theorems, we can conclude that a maximum size stable matching $m'$ under $\mathcal{M}'$ is of the size of the maximum size stable matching $m$ under $\mathcal{M}$ plus $|A''| + |A'''|$. Moreover, we have shown a reduction from the maximum size stable matching from $\mathcal{M}$ to $\mathcal{M'}$. Since for the latter instance each tie occur only at the end of each preferences list, is of length two, and no agent occur in more than one preferences list, we conclude that the SMTI problem is still hard under these settings.
\end{proof}

\begin{theorem}
\label{thm:WSM-special-case-hard}
It is NP-hard to compute a maximum weight stable matching, even when ties occur on one side only, each agent's order has at most one tie, each tie involves two agents, and every agent appears in the tie of at most one agent's order.
\end{theorem}

\begin{proof}
    The proof is by a simple reduction from SMTI, such that in the reduced instance the preferences and the ties are the same, and we complete incomplete lists in some arbitrary order. The utility function to maximize defined to by 1 for every two agents that appear in each other original preferences lists, and zero otherwise. It can be seen that an optimal solution the reduced instance have the same utility as the size of an optimal solution to the original instance, and vice versa.
\end{proof}


\thmNphSmallWorldsPrivate*

\begin{figure*}[t]
\centering
        \begin{tikzpicture}
        \pgfplotsset{set layers}
        \begin{axis}[
        width=0.5\textwidth,
        xmin=0, ymin=0,
        xmax=1.15, ymax=14, 
        tick label style={font=\small},
        xtick={0,0.2, 0.6, 1},
        xticklabels={0 ($\omega_1$), 0.2 ($\mu$), $q$, 1 ($\omega_2$)},
        ytick=\empty,
        yticklabels=\empty,
        xticklabel style={below right, xshift=-6},
        axis lines=left,
        axis line style=-{Latex[scale=1.2]},
        major tick length=0.8mm,
        clip=false,
        ]
          \addplot[red,densely dashed,semithick, domain=0:1] (x, 12.5) node[yshift=0,black,right,pos=1,font=\footnotesize]{$b_{i,j+2}$};
          \addplot[red,semithick, domain=0:1] (x, 10.3-1.5*x) node[yshift=2,black,right,pos=1,font=\footnotesize]{$b_{i,j+1}$};
          \addplot[red,semithick, domain=0:1] (x, 10+1.5*x) node[yshift=-2,black,right,pos=1,font=\footnotesize]{$b_{i,j}$};
          \addplot[red,densely dashed,semithick, domain=0:1] (x, 8) node[yshift=0,black,right,pos=1,font=\footnotesize]{$b_{i,j-1}$};
          \addplot[blue,densely dashed,semithick, domain=0:1] (x, 6) node[yshift=0,black,left,pos=0,font=\footnotesize]{$a_{i,j+1}$};
          \addplot[blue,densely dashed,semithick, domain=0:1] (x, 2) node[yshift=3,black,left,pos=0,font=\footnotesize]{$a_{i,j-1}$};
          \addplot[blue,densely dashed,semithick, domain=0:1] (x, 1.4) node[yshift=-2,black,left,pos=0,font=\footnotesize]{$a_{i,1}$};
          \addplot[blue,semithick, domain=0:1] (x, 5-5*x) node[yshift=-2,black,left,pos=0,font=\footnotesize]{$a_{i,j}$};
          \addplot[blue,densely dashed,semithick, domain=0:1] (x, 7) node[yshift=2,black,left,pos=0,font=\footnotesize]{$a'$};
          \addplot[dotted,semithick] coordinates {(0.2, 0) (0.2, 13)};
          \addplot[dotted,semithick] coordinates {(0.6, 0) (0.6, 7.5)};
          \addplot[dotted,semithick] coordinates {(1, 0) (1, 13)};
          \addplot[blue!50, draw=none, mark=*, mark options={fill=blue,draw opacity=0,scale=0.5}] coordinates {(0,0) (0,1.4) (0,2) (0,5) (0,6) (0,7)
          (0.2,1.4) (0.2,2) (0.2,4) (0.2,6) (0.2,7)
          (1,0) (1,1.4) (1,2) (1,6) (1,7)
          (0.6,1.4) (0.6,2) (0.6,6) (0.6,7)};
          \addplot[red!50, draw=none, mark=*, mark options={fill=red,draw opacity=0,scale=0.5}] coordinates {(0,8) (0,10) (0,10.3) (0,12.5)
          (0.2,8) (0.2,10.3) (0.2,10) (0.2,12.5)
          (1,8) (1,11.5) (1,8.8) (1,12.5)};
          \draw [decorate, decoration={calligraphic brace,amplitude=10pt}, thick, pen colour={black}, black] (-20,0) -- (-20,70) node [midway, anchor=east, xshift=-2, outer sep=10pt]{$v_{b_i}(\cdot \given p)$};
          \draw [decorate, decoration={calligraphic brace,amplitude=8pt}, thick, pen colour={black}, black] (118,125) -- (118,80) node [midway, anchor=west, xshift=2, outer sep=10pt]{$v_{a_i}(\cdot \given p)$};
        \end{axis}
        \end{tikzpicture}
        \caption{Agents' values in the proof of \Cref{thm:nph-small-worlds-private}. The x-axis represent $p(\omega_2)$. The values may not be precisely the ones defined in the proof; only the relative orders at $p(\omega_1) = 0$, $0.2$, $q$, and $1$ within each group ($v_{a_i}$ or $v_{b_i}$) are important.} 
        \label{fig:private-persuasion-hard-reduction}
\end{figure*}
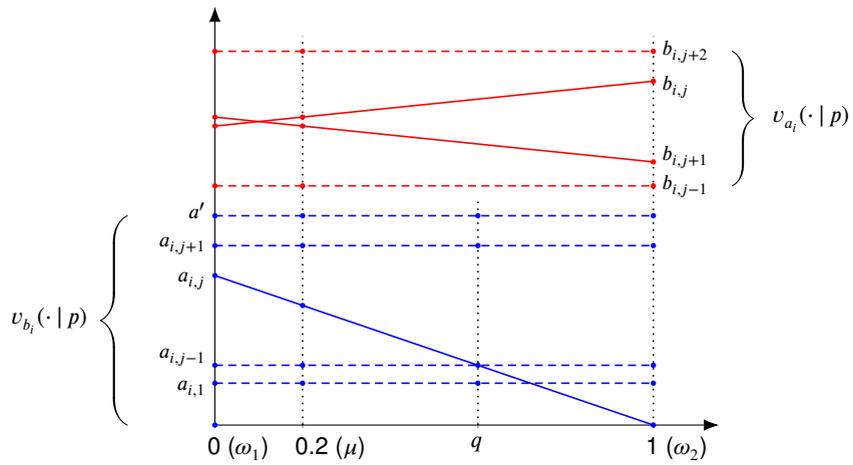

\begin{proof}
We prove via a reduction from the special case of the WSM problem in \Cref{thm:WSM-special-case-hard}, where there is at most one tie in the preference order of each agent $a \in A$ and no tie in the preference order of any agent $b \in B$, each tie involves two agents, and every $b\in B$ appears in at most one tie.
Pick an arbitrary WSM instance in this special case. Suppose $A = \{a_1, \dots, a_n\}$ and $B = \{b_1, \dots, b_n\}$. And we write the preference order $\preceq_{a_i}$ of each agent $a_i \in A$ in this instance as 
\[
b_{i1} \prec_{a_i} b_{i2} \prec_{a_i} \prec_{a_i} \dots b_{ij} \sim_{a_i} b_{i,j+1} \prec_{a_i} \dots \prec_{a_i} b_{in},
\]
where $b_{ij}$ and $b_{i,j+1}$ are in a tie. Similarly, the preference order $\prec_{b_i}$ of each $b_i \in B$ is $a_{i1} \prec_{b_i} \dots \prec_{b_i} a_{in}$ (which by assumption involves no tie).

\paragraph{Reduction}
We reduce the instance to a private persuasion instance as follows.
Let $\Omega = \{\omega_1, \omega_2\}$ and $\mu = (0.8, 0.2)$.
Let there be $n+1$ agents on each side: $A^* = A \cup \{a'\}$ and $B^* = B \cup \{b'\}$.

The agents' values are set as follows, as also illustrated in Figure~\ref{fig:private-persuasion-hard-reduction}.
\begin{itemize}
\item 
For each agent $a_i$:
    \begin{itemize}
    \item  
    For every $b_{ij}$ not in the tie in $\preceq_{a_i}$, let 
    \[
    v_{a_i}(b_{ij} \given \omega_1) = v_{a_i}(b_{ij} \given \omega_2) = j.
    \]
    
    \item 
    For every $b_{ij}$ and $b_{i,j+1}$ in the tie, let 
    \begin{align*}
    &v_{a_i}(b_{ij} \given \omega_1) = j + 0.4, \\
    &v_{a_i}(b_{ij} \given \omega_2) = j + 1.4, \\
    &v_{a_i}(b_{i,j+1} \given \omega_1) = j + 0.6, \\
    \text{and }\quad
    &v_{a_i}(b_{i,j+1} \given \omega_2) = j - 0.4.
    \end{align*}
    The numbers ensure $p < 0.2 = \mu(\omega_2)$ for the point $p$ where $v_{a_i}(b_{ij} \given p) = v_{a_i}(b_{i,j+1} \given p)$. Meanwhile, $v_{a_i}( \cdot \given \omega_1)$ and $v_{a_i}( \cdot \given \omega_2)$ induce two strict orders that are exactly the two strict orders resulting from breaking the tie in $\preceq_{a_i}$.
    
    \item 
    Let her value for $b'$ be $0$ in every world.
    \end{itemize}

\item 
For each agent $b_i$:
    \begin{itemize}
    \item 
    Suppose that $b_i$ appears in the tie in $a_{ij}$'s order (by assumption there is at most one such agent in $A$), let \begin{align*}
    &v_{b_i}(a_{ij} \given \omega_1) = 2, \\
    \text{and }\quad
    &v_{b_i}(a_{ij} \given \omega_2) = 0.
    \end{align*}

    \item 
    Let her value for $a'$ be 
    \begin{align*}
    &v_{b_i}(a' \given \omega_1) = 3.9, \\
    \text{and }\quad
    &v_{b_i}(a' \given \omega_2) = 3.9.
    \end{align*}
    
    \item 
    For $a_{i1},\dots, a_{i,j-1}$, let $b_i$'s value for each of them be equal in the two worlds, and let the values be distributed in the interval $(0.5, 1)$ in any arbitrary way such that $v_{b_i}(a_{i1} \given \omega_1) < v_{b_i}(a_{i2} \given \omega_1) < \dots < v_{b_i}(a_{i,j-1} \given \omega_1)$.

    \item 
    Similarly, let $b_i$'s value for each of $a_{i,j+1},\dots, a_{in}$ be the equal in the two worlds, and let the values be distributed in the interval $(2, 3)$ in any arbitrary way such that \\ $v_{b_i}(a_{i,j+1} \given \omega_1) < v_{b_i}(a_{i,j+2} \given \omega_1) < \dots < v_{b_i}(a_{in} \given \omega_1)$.
    \end{itemize}

\item 
For agent $a'$, let her value for $b'$ be $0$ in every world, and her value for every $b \in B$ be $-1$ in $\omega_1$ and $1$ in $\omega_2$. 
This gives $v_{a'}(b' \given p) = v_{a'}(b \given p)$ for all $b \in B$ at $p = (0.5, 0.5)$. Hence, $a'$ prefers $b'$ when $p(\omega_2) < 0.5$ and prefers other agents when $p(\omega_2) > 0.5$.

\item 
For agent $b'$, let her values be $1$ for $a'$ and $0$ for any other $b\in B$ in both worlds.

\end{itemize}

Finally, we let the principal's utility for each pair $(a,b) \in A\times B$ be the same as the weight of $(a,b)$ in the WSM instance. Let her utility be $0$ for $(a', b')$, and $-n$ for any other pair. Without loss of generality, we assume that all weights in the WSM instance are in $[0,1]$. 

\paragraph{Correctness of the Reduction}

We show that, for any arbitrary optimal policy in the above instance, every matching used by this policy must involve an optimal matching in the WSM instance. Hence, an efficient algorithm that computes an optimal policy also solves the WSM problem and the reduction is complete.

First, we show that the principal's optimal utility in the private persuasion (PP) instance is at least as much as the optimal utility in the WSM instance, i.e.,  $\mathrm{OPT}_{\mathrm{WSM}} \le \mathrm{OPT}_\mathrm{PP}$.
To see this, suppose that $M^*$ is optimal in the WSM instance. By definition, $M^*$ is stable with respect to the preferences $\preceq$ given in the WSM instance. Hence, it must also be stable with respect to some strict profile $\prec^*$ that results from resolving ties in $\preceq$ (for the same reason as Claim~(1) in the proof of \Cref{lmm:Cpp}).
Consider the following private policy $\sigma$:
\begin{itemize}
\item 
For every agent $a \in A$ whose order at $\mu$ is different from $\prec_a^*$, the policy always reveals truth (i.e., it signals them their preference order at the realized world).
Additionally, if $M^*(a)$ is an agent in the tie in $\preceq_a$, then the policy also always reveals truth to {\em the other} agent in the tie.

\item 
For all the other agents, the policy always reveals no information (i.e., it signals them their preference order at $\mu$). 
\end{itemize}
In addition to the above signals, the policy always selects the matching $\widetilde{M}^* = M^* \cup \{(a', b')\}$ in every meta-signal.

It can be verified that this private policy is stable given that $M^*$ is stable in the original instance.
Indeed, when it is world $\omega_1$, the preferences induced among agents in $A\cup B$ are exactly the same as $\prec^*$.
When it is world $\omega_2$, each $a_i \in A$ whose order is different from $\prec^*$ will derive a different order, in which $b_{ij}$ and $b_{i,j+1}$ are switched. 
Now, if $a_i$ is matched to neither $b_{ij}$ nor $b_{i,j+1}$ in $M^*$, this will not change the stability.
If $a_i$ is matched to one of them, say $b_{ij}$, the change of the order between $b_{ij}$ nor $b_{i,j+1}$ may cause $a_i$ to prefer $b_{i,j+1}$ to $b_{ij}$. Nevertheless, this will not make $\widetilde{M}^*$ unstable because $b_{i,j+1}$ will also be informed about the world $\omega_2$, in which case $a_i$ will drop to the bottom of $b_{i,j+1}$'s order according to the construction. 

Hence, we can conclude that 
\[
\mathrm{OPT}_{\mathrm{WSM}} \le \mathrm{OPT}_\mathrm{PP}.
\]
This also means that every optimal policy $\sigma^*$ of the PP instance must use at least one matching where $a'$ and $b'$ are matched together (otherwise, the policy would always select matchings with negative utilities, given the large penalty $-2n$ for not matching $a'$ and $b'$ together).
Pick an arbitrary matching $\widetilde{M} = M \cup \{(a',b')\}$ used by $\sigma^*$.
We further argue that $M$ must be an optimal solution to the WSM instance.

Suppose for the sake of contradiction that $M$ is not an optimal solution.
This means one of the following cases:
either $M$ is a feasible solution but $u(M) < u(M^*)$, or $M$ is infeasible (i.e., unstable).
In the first case, there must be another matching $\widetilde{M}' = M' \cup \{(a',b')\}$ used by the policy such that $u(M') > u(M^*)$: otherwise, we would have $\mathrm{OPT}_{\mathrm{PP}} = u(\sigma^*) < u(M^*) = \mathrm{OPT}_{\mathrm{WSM}}$, which contradicts our conclusion above.
By assumption $u(M^*)$ is optimal in the WSM instance, so $u(M') > u(M^*)$ implies that $M'$ is not a feasible solution to the WSM instance, which is the same as the second case. We analyze the second case next, i.e., $\sigma^*$ uses a matching $\widetilde{M} = M \cup \{(a',b')\}$ where $M$ is unstable in the WSM instance.


Let $(a,b)$ be a pair that blocks $M$ from being stable in the WSM instance,
i.e., we have 
\begin{equation}
\label{eq:a-b-block-M}
M(a) \prec_a b
\quad\text{and}\quad
M(b) \prec_b a
\end{equation}
(i.e., $M(a) \preceq_a b$ but $b {\not\preceq}_a M(a)$, and similarly for the second relation).
Pick an arbitrary meta-signal $(g, \widetilde{M})$ in the support of $\sigma^*$ and consider the posterior belief $p_x := \Pro( \cdot \given g_x, M)$ it induces for each agent $x$.
Since $\sigma^*$ is optimal, by definition $\widetilde{M}$ must be stable with respect to $p_a$, which means 
\[
v_a(M(a) \given p_a) \ge v_a(b \given p_a)
\quad\text{or}\quad
v_b(M(b) \given p_b) \ge v_b(a \given p_b).
\]
Since we have $M(a) \prec_a b$,
the first inequality above cannot hold: by construction $v_a(M(a) \given p_a)$ and $v_a(b \given p_a)$ would never be equal unless $M(a) \sim b$ in $\preceq_a$ (see \Cref{fig:private-persuasion-hard-reduction}).

Now consider the second inequality $v_b(M(b) \given p_b) \ge v_b(a \given p_b)$. This implies a different preference order from the one at $\mu$. So according to the value function of $b$, it must be that $p_b(\omega_2) \ge q$, where $q$ is the point at which $v_b( x \given q) = v_b( y \given q)$ (see \Cref{fig:private-persuasion-hard-reduction}), agent $x \in A$ is the one whose tie involves $b$, and $y$ is the agent immediately before $x$ in $b$'s order. 
By construction, $v_{b}(y \given p) < 1$ for all $p$, so we have 
$q > 0.5$.
Using the following inequality
\[
\Pro(\omega_2 \given \widetilde{M}) \le 0.5
\]
(see Claim~(1) below)
we get that 
\[
p_b(\omega_2) \ge q > 0.5 \ge \Pro(\omega_2 \given \widetilde{M}).
\]
Recall that $p_b(\omega_2) := \Pro(\omega_2 \given g_b, M)$, and we can write $\Pro(\omega_2 \given \widetilde{M})$ as a convex combination involving $\Pro(\omega_2 \given g_b, \widetilde{M})$, i.e., 
\[
\Pro(\omega_2 \given \widetilde{M}) = \sum_{g'_b} \Pro(\omega_2 \given g'_b, \widetilde{M}) \cdot \Pro(g'_b \given \widetilde{M}).
\]
Hence, the fact that $\Pro(\omega_2 \given g_b, \widetilde{M}) > \Pro(\omega_2 \given \widetilde{M})$ means that there exists another meta-signal $g'_b$ such that 
\[
\Pro(\omega_2 \given g'_b, \widetilde{M})
<
\Pro(\omega_2 \given \widetilde{M}).
\]
Consequently, $\Pro(\omega_2 \given g'_b, \widetilde{M}) < 0.5$, so the preference profile under $\Pro(\cdot \given g'_b, \widetilde{M})$ is the same as $\preceq$ in the WSM instance. In this case, $a$ and $b$ would block $\widetilde{M}$ from being stable according to \Cref{eq:a-b-block-M}, i.e., 
\[
\widetilde{M}(a) = M(a) \prec_a b 
\quad\text{and}\quad
\widetilde{M}(b) = M(b) \prec_b a.
\]
This contradicts the assumption that $\sigma^*$ is an optimal policy (and hence stable).

\paragraph{Claim (1).}
$\Pro(\omega_2 \given \widetilde{M}) \le 0.5$.

\begin{proof}[Proof of Claim~(1)]
According to the value function of $a'$, when $a'$ is matched to $b'$, the posterior probability $\Pro(\omega_2 \given g_{a'}, M)$ agent $a'$ derives upon receiving any meta-signal $(g_{a'}, M)$ is at most $0.5$, as otherwise $a'$ will prefer agents $b \in B$ to $b'$ while being the favorite of these agents. 
It follows that
$\Pro(\omega_2 \given \widetilde{M}) 
= \sum_{g_{a'}} \Pro(\omega_2 \given g_{a'}, M) \cdot \Pro(g_{a'} \given \widetilde{M})
\le 0.5$.
\end{proof}



\paragraph{Smoothed Analysis}
The relative orders of the values remain the same at $p(\omega_1) = 0$, $\mu(\omega_2)$, $q$, and $1$ under small perturbations to the parameters, so the set of hard instances has a nonzero measure. 
This completes the proof.
\end{proof}

\end{document}